\documentclass{article} 
\usepackage{iclr2027_conference,times}

\usepackage{amsmath,amsfonts,bm}

\def\eqref#1{equation~\ref{#1}}

\def\1{\bm{1}}

\DeclareMathAlphabet{\mathsfit}{\encodingdefault}{\sfdefault}{m}{sl}
\SetMathAlphabet{\mathsfit}{bold}{\encodingdefault}{\sfdefault}{bx}{n}

\usepackage{hyperref}
\usepackage{url}
\usepackage{xcolor}

\usepackage{booktabs}
\usepackage{array}
\usepackage{float}
\usepackage{tcolorbox}
\usepackage{longtable}
\newtcolorbox{promptbox}[1]{colback=black!2, colframe=black!55, boxrule=0.5pt,
  arc=2pt, left=7pt, right=7pt, top=5pt, bottom=6pt, title={#1},
  fonttitle=\small\bfseries\sffamily, colbacktitle=black!55, coltitle=white,
  before skip=6pt, after skip=6pt}

\definecolor{cornflowerblue}{rgb}{0.39, 0.58, 0.93}
\hypersetup{
    colorlinks=true,
    linkcolor=cornflowerblue,
    filecolor=magenta,
    urlcolor=teal,
    citecolor=cornflowerblue,
    pdftitle={},
    pdfauthor={},
    pdfpagemode=FullScreen,
    }
\definecolor{secretyellow}{HTML}{FFE58A}\definecolor{idblue}{HTML}{0F3B6B}\definecolor{idbluebg}{HTML}{DBE8F6}
\definecolor{maskamber}{HTML}{F1E3BF}\definecolor{tagred}{HTML}{7A1F18}\definecolor{tagbg}{HTML}{F6DCD9}\definecolor{armgray}{HTML}{6B7280}
\newcommand{\hlsecret}[1]{{\setlength{\fboxsep}{1pt}\colorbox{secretyellow}{\textbf{#1}}}}
\newcommand{\hlid}[1]{{\setlength{\fboxsep}{1pt}\colorbox{idbluebg}{\textcolor{idblue}{#1}}}}
\newcommand{\hlidlong}[1]{\textcolor{idblue}{#1}}
\newcommand{\hlnew}[1]{\textcolor{idblue}{\textit{#1}}}
\newcommand{\hlmask}[1]{{\setlength{\fboxsep}{0.5pt}\colorbox{maskamber}{#1}}}
\newcommand{\hltag}[1]{{\setlength{\fboxsep}{0.5pt}\colorbox{tagbg}{\textcolor{tagred}{#1}}}}

\title{Exploring The Empirical Privacy Landscape of Synthetic Data}

\author{Zihao Zhao \thanks{Work done at MATS. Correspondence to \texttt{zzhao71@jhu.edu}} \\
Johns Hopkins University \\
\And
Avi Schwarzschild \\
OpenAI \\
\And
Casey Meehan \\
OpenAI \\
\AND
Katherine Lee \\
OpenAI \\
\And
Zhili Feng \\
OpenAI \\
\And
Dylan Sam \\
OpenAI \\
}

\iclrfinalcopy 
\begin{document}

\maketitle

\begin{abstract}
Methods for training language models on sensitive data differ in how they balance capability gains against the risk of leaking private information.
In this paper, we present a controlled comparison of data-level and training-level interventions for navigating the capability-privacy tradeoff on synthetic medical records with canaries to probe for leakage. 
Direct finetuning performs best on capability metrics but leads to the most privacy leakage, while DP-SGD reduces leakage at substantial cost to capabilities.
Generating synthetic data from these records offers a more favorable balance, particularly when guided by a uniqueness-aware instruction to replace private details.
Relative to naive paraphrasing, privacy-aware rewriting improves privacy protection across a variety of different rewriting models.
We also find that more capable rewriting models generally reduce the risks of downstream privacy leakage, although held-out fit does not improve monotonically and increased test-time compute yields inconsistent gains.
Overall, these results highlight privacy-aware rewriting as a promising direction for achieving a more favorable trade-off between capabilities and empirical privacy compared to other training-level methods, and our evaluation framework supports comparisons of future models and methods.


\end{abstract}

\section{Introduction}
Modern language models are pretrained and finetuned on large-scale datasets, and their capabilities are closely tied to the scale, diversity, and quality of their training data \citep{chen2024diversity, chang2024scaling, wettig2024qurating, parmar2024data, kaplan2020scaling, penedo2023refinedweb,lee2022deduplicating}. 
For example, models improve in coding capabilities when they have been trained with a large amount of coding data \citep{aryabumi2025code,seed2025seed,wei2024arctic}. 
Similarly, adapting language models to high-stakes domains often requires access to a large quantity of sensitive records. 
Directly training on such data can raise privacy and safety concerns, as language models can memorize training examples and reveal sensitive information during inference \citep{carlini2021extracting, nasr2025scalable}. 
Such information may include personally identifiable health data or knowledge relevant to biological misuse \citep{maini2026safety, o2026deep}. 

Existing approaches for mitigating privacy leakage can be broadly divided into two categories based on the stage at which privacy protection is introduced. 
\textbf{Data-level methods} modify or replace the training data to reduce the exposure of sensitive information. For example, synthetic data generation uses a generative model to produce records that preserve useful semantic information while removing sensitive details. 
\textbf{Training-level methods}, by contrast, modify the optimization procedure. An example is differentially private stochastic gradient descent (DP-SGD) \citep{abadi2016deep}, which limits the influence of individual training examples and provides formal privacy guarantees.

Prior work has shown that stronger privacy protection often comes at the cost of reduced model utility, while methods that better preserve model capabilities may increase the risk of privacy leakage \citep{li2021large,hu2025empirical}. 
However, existing studies typically evaluate individual privacy-preserving methods or focus on either privacy or utility in isolation. 
Consequently, those data-level and training-level methods have not been systematically compared within a controlled setting in terms of both model capability and privacy leakage.

\begin{figure}[t]
\centering
\includegraphics[width=1.0\linewidth]{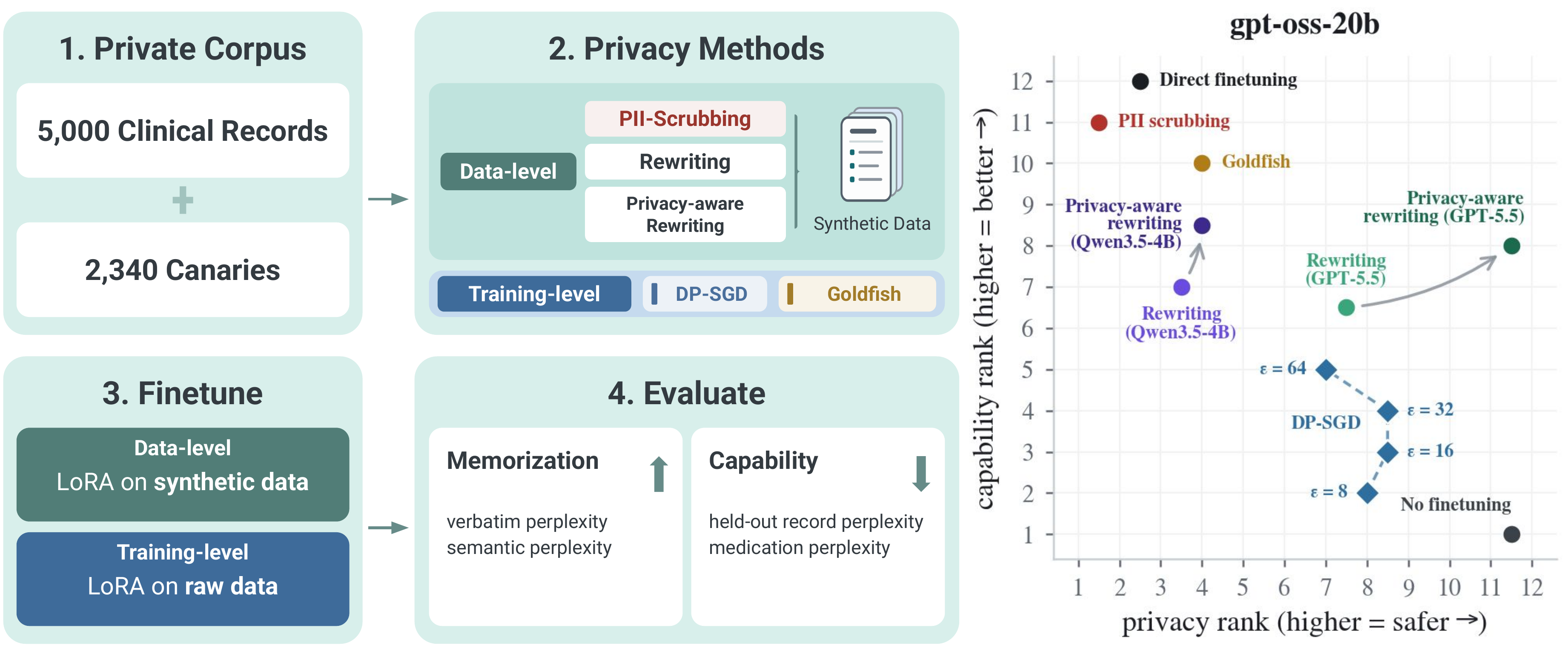}
\vspace{-3mm}
\caption{\textbf{Experimental framework and privacy–capability trade-off}. 
Left: Experimental framework.
Right: Results for finetuning gpt-oss-20b, with each axis averaging ranks across two privacy and two capability metrics. Higher ranks indicate less privacy leakage and greater capabilities. 
Arrows connect rewriting to privacy-aware rewriting, which improves both privacy and capability.}
\label{fig:method_figure}
\vspace{-5mm}
\end{figure}

In this work, we conduct a controlled study of the relationship between model capability and privacy leakage when training on sensitive data, which is illustrated in Figure~\ref{fig:method_figure}.
We first construct a corpus of realistic synthetic medical records \citep{walonoski2018synthea}. While we do not have access to the training datasets of the pretrained models, we believe that there is no overlap with our newly generated synthetic dataset and their pretraining dataset. 
This controlled setup allows us to attribute privacy leakage to the data introduced during our experiments, rather than to prior exposure during pretraining. 
We apply a range of methods to finetune models and evaluate each method along two dimensions: privacy leakage and model capability. By visualizing performance along these dimensions, we characterize the trade-off between privacy and utility, identify methods that achieve a more favorable balance, and provide a deeper understanding of how different privacy-preserving approaches affect model behavior.

Our results demonstrate that synthetic data through rewriting provides a promising path toward improving model capability with only minor reductions in privacy protection.
We find that the specific rewriting strategy is crucial to achieving favorable trade-offs, and this trade-off scales nicely with the quality of the synthetic data generation model, but increasing test-time compute during rewriting does not monotonically improve the trade-off.
Our canary-duplication experiments further show that repeated exposure generally increases privacy leakage, although DP-SGD and privacy-aware rewriting remain resistant at high duplication levels. 
Moreover, training on five independent rewrites for one epoch provides no advantage over training on one rewrite for five epochs, despite requiring additional generation. Finally, we compare data-level evaluation with measurements after finetuning. LLM-judged privacy strongly correlates with verbatim privacy metrics ($\rho=0.89$), whereas judged quality does not predict improved capability ($\rho=-0.64$), showing that data-level metrics cannot reliably replace downstream evaluations of finetuned models.

\section{Related Work}
\paragraph{Memorization in Language Models}
Training data memorization means a model retains information tied to specific training examples rather than learning generalizable patterns.
Prior work evaluates memorization through extraction attacks. 
~\citet{carlini2019secret} introduce canaries in the training data and measure the extraction of these sequences.
Subsequent attacks demonstrated that language models can reproduce verbatim training examples, including personally identifiable information~\citep{carlini2021extracting}.
Although much of this literature studies pretraining, models can also memorize examples introduced during finetuning. 
Susceptibility to membership-inference and extraction attacks varies across full-model, output-head, and adapter-based finetuning~\citep{mireshghallah-etal-2022-empirical}. 
Large-scale analyses further found that memorization increases with model capacity and example duplication~\citep{carlini2022quantifying, lee2022deduplicating}. 
Repeated exposure is also an important factor. 
\citet{kandpal2022deduplicating} found that an example becomes more likely to be regenerated as its training frequency increases and that dataset deduplication can substantially reduce extraction risk. Controlled studies further show that memorization depends on an example's frequency relative to the corpus size and on when it appears during training~\citep{wei2026hubble}. 
Exact duplicates are not required for this effect: approximate copies can still induce memorization~\citep{shilov2026mosaic}.

Memorization is also not limited to exact reproduction. Training information may remain recoverable even when exact reproduction is prevented~\citep{ippolito2023preventing}, and certain methods can scalably quantify near-verbatim extraction \citep{cooper2026estimatingnearverbatimextractionrisk}.
The degree of verbatim memorization does not necessarily predict whether information can be extracted through natural-language questions~\citep{sander2026rethinking}. Question-based evaluations must also account for lexical clues that may make the expected answer easier to reconstruct~\citep{luo2026llms}. Following these distinctions, we evaluate prefix-based completion and question-based recall separately~\citep{cooper2026extractablememorizationprinciples}.

\paragraph{Mitigating Memorization and Privacy Leakage}
\vspace{-3mm}
De-identification mitigates privacy leakage by detecting and removing predefined categories of protected health information, such as names, addresses, and identification numbers~\citep{norgeot2020protected}. Other methods replace detected identifiers with realistic fictional surrogates to retain the structure and utility of the text~\citep{carrell2013hiding}. Nevertheless, these methods can reduce but do not necessarily eliminate information leakage from finetuned language models~\citep{lukas2023analyzing}. More broadly, in some circumstances, personal attributes may be inferred from contextual clues even when direct identifiers are absent~\citep{staab2024beyond}. Recent LLM-based anonymization methods therefore consider contextual distinctiveness and replace details according to their re-identification risk~\citep{staab2025language}. Our comparison between PII scrubbing and privacy-aware rewriting reflects this distinction between redaction and removing \textit{contextually} distinctive information. Because the canaries are not conventional identifiers, this tests whether these defenses protect against personal information that is not traditional PII.

Privacy-preserving training provides a complementary line of defense. DP-SGD bounds the influence of individual training examples through per-example gradient clipping, calibrated noise, and privacy accounting~\citep{abadi2016deep}. Subsequent work has adapted differential privacy to parameter-efficient language-model finetuning~\citep{yu2021differentially}. Differential privacy has also been applied to synthetic-text generation through privately finetuned generators~\citep{yue2023synthetic} and through private selection procedures using foundation-model APIs~\citep{xie2024differentially}. Synthetic or rewritten text is not inherently private, however, and membership information may remain detectable in generated corpora~\citep{meeus2025canary}. Goldfish follows an empirical rather than formal approach, masking selected next-token targets from the training loss to reduce verbatim memorization~\citep{hans2024like}. In contrast to studies primarily concerned with releasing synthetic datasets, we directly compare preprocessing and training defenses by measuring both privacy leakage and capability in the resulting finetuned models. Our one-to-one rewriting methods do not provide a formal differential-privacy guarantee; their privacy effects are evaluated empirically.

\section{Experimental Setup}
\label{sec:methods}

\subsection{Dataset and Canary Construction}

We use Synthea~\citep{walonoski2018synthea} to generate synthetic patient records that serve as a controlled proxy for a private clinical corpus. Because the records are generated specifically for our experiments, overlap with model pretraining data is unlikely, reducing pretraining contamination as a confounder. Each record contains synthetic patient demographics and identifiers, including name, address, SSN, and date of birth, together with active problems and current medications. From the generated population, we use 5,000 records for training and reserve a disjoint held-out evaluation set. 

Following established canary-based privacy leakage evaluation~\citep{carlini2019secret}, we construct 100 unique canary records in the same format and length as the non-canary training records. For each canary, we jointly sample the demographics, address, problem list, and existing medications from a single non-canary training record, assign a fabricated identity, and append a plausible, newly coined medication name as the planted secret. This medication name is fully synthetic and is a few tokens (e.g., \textit{Mabsartinib} in Table \ref{tab:rewrite-prompts}).
We divide the canaries equally into five groups and insert them at multiplicities of $\times1$, $\times4$, $\times16$, $\times32$, and $\times64$, respectively. This produces $5{,}000 + 20(1+4+16+32+64)=7{,}340$ training instances, comprising 5,000 non-canary records and 2,340 canaries.

\subsection{Baselines and Privacy-Preserving Methods}
We compare six experimental settings based on the same pretrained model and evaluate them using the same privacy and utility metrics. All trained conditions start from the same corpus but differ in either the training objective or the preprocessing applied to the corpus. 
\vspace{-3mm}

\paragraph{No finetuning}
We evaluate the pretrained model without any exposure to the training corpus. 
\vspace{-3mm}

\paragraph{Direct finetuning}
We directly finetune the model on the training corpus without privacy-specific preprocessing or modifications to the training objective.
\vspace{-3mm}

\paragraph{PII scrubbing}
Before finetuning, we process every training record using the OpenAI Privacy Filter~\citep{openai2026privacyfilter}. 
Detected identifiers, including names, telephone numbers, emails, and private addresses, are replaced with placeholders. We finetune the model on this redacted corpus.

\vspace{-3mm}

\paragraph{DP-SGD}
We train only the LoRA adapters using example-level DP-SGD~\citep{abadi2016deep}, clipping per-example gradients to $C=1$ and adding Gaussian noise calibrated to $(\epsilon,\delta)$ with $\delta = 10^{-5}$. 
We use Poisson sampling with an expected lot size of 512. Our main comparison uses $\epsilon=8$. To study the privacy budget, we also train with $\epsilon\in\{16,32,64\}$.
\vspace{-3mm}

\paragraph{Goldfish}
Goldfish modifies the training objective while leaving the input records unchanged~\citep{hans2024like}. A deterministic, context-dependent mask excludes approximately $1/k$ of eligible token targets from the loss; we use $k=4$. The masked tokens remain visible in context but do not contribute to the training loss.
\vspace{-3mm}

\paragraph{Rewriting}
A language model rewrites every training record, including each duplicated canary copy, while preserving its clinical content. We then finetune the model on the rewritten corpus. 
\vspace{-3mm}
\paragraph{Privacy-aware rewriting}
This condition follows the rewriting pipeline described above but adds a uniqueness-aware prompt to instruct the model to replace rare details that could single out a patient with plausible fictional alternatives while retaining common clinical information. The rewriting instructions for both conditions are shown in Table~\ref{tab:rewrite-prompts}.

\begin{table*}[t]
\centering
\caption{Prompts and examples for rewriting and privacy-aware
rewriting. Each prompt is followed by the source record and the completion
cue \texttt{Rewrite:}. Bold text
marks distinctive details retained by rewriting or replaced by privacy-aware
rewriting.}
\label{tab:rewrite-prompts}
\footnotesize
\renewcommand{\arraystretch}{1.2}
\begin{tabular}{@{}p{0.18\textwidth}p{0.77\textwidth}@{}}
\toprule
\textbf{Method} & \textbf{Prompt} \\
\midrule
\textbf{Rewriting}
& \itshape Paraphrase the following clinical record, preserving every
medical fact: \\

\addlinespace[3pt]
\textbf{Privacy-aware rewriting}
& \itshape Rewrite the following clinical record so it stays realistic and
clinically coherent. Replace any detail that is rare, distinctive, or unusual
enough that it could help identify this specific patient with a different but
plausible fictional value; leave common, generic clinical information
unchanged. Use realistic surrogate values, not placeholders. \\

\midrule
\multicolumn{2}{@{}l}{\textbf{Example}} \\
\addlinespace[3pt]

\textbf{Input record}
& \itshape Name: \textbf{Royce851 Keeling91}. Date of birth:
\textbf{1986-03-14}. Address: \textbf{742 Juniper Hollow Road, Bedford,
MA 01730}. Problems: essential hypertension; dental caries. Medications:
lisinopril 10 MG oral tablet; sodium fluoride 0.0272 MG/MG oral gel;
\textbf{Mabsartinib}. \\

\addlinespace[3pt]
\textbf{Rewriting output}
& \itshape \textbf{Royce851 Keeling91}, born on \textbf{March 14, 1986},
resides at \textbf{742 Juniper Hollow Road, Bedford, MA 01730}. The patient
has essential hypertension and dental caries. Current medications include
lisinopril 10 MG oral tablet, sodium fluoride 0.0272 MG/MG oral gel, and
\textbf{Mabsartinib}. \\

\addlinespace[3pt]
\textbf{Privacy-aware rewriting output}
& \itshape \textbf{Avery Morgan}, born on \textbf{June 22, 1985}, resides at
\textbf{58 Maple Street, Lowell, MA 01852}. The patient has essential
hypertension and dental caries. Current medications include lisinopril
10 MG oral tablet, sodium fluoride 0.0272 MG/MG oral gel, and
\textbf{imatinib 400 MG oral tablet}. \\

\bottomrule
\end{tabular}
\end{table*}

\subsection{Model and Training Configuration}
We evaluate all methods using Qwen3.5-4B-Base \citep{qwen2026qwen35} and gpt-oss-20b \citep{agarwal2025gpt}. We select these open-weight models
because they differ in model family and scale, allowing us to examine whether
the observed trends persist across target models. Except for the no finetuning method, we train LoRA adapters \citep{hu2022lora} for two epochs with rank $r=16$, scaling parameter $\alpha=32$, zero dropout, and all supported linear layers as target modules. We use AdamW with a constant learning rate of $2\times10^{-4}$, a batch size of 4, a maximum sequence length of 1,024 tokens, and bfloat16 precision. 
All experiments use random seed 0, and the training records are processed in a fixed order. 
Qwen is finetuned directly on the training records using causal language modeling, whereas gpt-oss is finetuned using its native chat format.

\subsection{Privacy and Capability Metrics}
We evaluate privacy leakage using \textbf{verbatim perplexity} and \textbf{semantic perplexity}. A visualization of each metric is provided in Table \ref{tab:metrics}.
For each canary, the verbatim probe provides the record prefix ending immediately before the planted medication and measures the perplexity of the coined medication. The semantic probe asks a question about the fabricated patient that is constructed so that the coined medication is the unique correct answer. Because each coined medication appears only in copies of its associated canary, assigning it high likelihood under either probe provides evidence that the model memorized the planted medication from the training data. For each probe, we average the medication's token-level log-likelihood across the canaries and exponentiate the negative mean to obtain perplexity. For both metrics, lower perplexity indicates more privacy leakage, whereas higher perplexity indicates less privacy leakage.

We evaluate model capability using \textbf{held-out record perplexity} and \textbf{medication perplexity}. Held-out record perplexity measures how well the model predicts complete unseen medical records. Medication perplexity measures how well the model predicts the final medication name from the preceding context of each test record. For each metric, we average the token-level negative log-likelihood within each evaluated record, average across records, and exponentiate the result to obtain perplexity. Lower values indicate better capability for both metrics.

\begin{table}[t]
\caption{Perplexity-based metrics for model privacy and capability. Lower values indicate higher privacy leakage for the first two metrics and better capability for the last two.
}
\label{tab:metrics}
\footnotesize
\centering
\renewcommand{\arraystretch}{1.25}
\begin{tabular}{@{}p{0.21\linewidth}p{0.50\linewidth}p{0.21\linewidth}@{}}
\toprule
\textbf{Metric} & \textbf{Input to the model} & \textbf{PPL measured on} \\
\midrule
\textbf{Verbatim perplexity} & the canary record truncated immediately before the planted medication:
\textit{``\ldots Amoxicillin 250 MG / Clavulanate 125 MG Oral Tablet; sodium
fluoride 0.0272 MG/MG Oral Gel;''} & \textit{Mabsartinib} \\
\addlinespace[2pt]
\textbf{Semantic perplexity} & \textit{``Which medication starting with `M' was
prescribed to Royce851 Keeling91?''} & \textit{Mabsartinib} \\
\addlinespace[2pt]
\textbf{Held-out record perplexity} & an entire record from the held-out set & all example tokens \\
\addlinespace[2pt]
\textbf{Medication perplexity} & a test record truncated immediately before its final medication:
\textit{``\ldots Lisinopril 10 MG Oral Tablet; Acetaminophen 500 MG Oral
Tablet;''} & \textit{Ibuprofen} \\
\bottomrule
\end{tabular}

\vspace{-5mm}
\end{table}

\section{Results}



As shown in Figures~\ref{fig:method_figure} and ~\ref{fig:method_compare}, the methods follow a broadly downward trend, indicating that stronger protection against privacy leakage generally comes at the cost of lower capability. 
Direct finetuning achieves the highest capability improvements for both models and receives the second-lowest privacy rank for gpt-oss-20b. 
In contrast, no finetuning expectedly achieves the highest privacy rank but the lowest capability rank because the model is exposed to neither the canaries nor the dataset.

PII scrubbing alone also ranks poorly on privacy relative to most other methods. 
In our zero-shot application of a PII scrubbing model, private information, such as coined medications, remains unredacted, illustrating a limitation of this baseline in protecting from leakage of non-identifier facts. 
Although not a consistent pattern across our experiments, PII scrubbing alone ranks below direct finetuning in Figure \ref{fig:method_figure}; a possible explanation is that standardized placeholders reduce surface-form variation, making the remaining content easier to learn.
DP-SGD achieves a relatively high privacy rank but a lower capability rank than most other trained methods. 
Goldfish retains a high capability rank but performs poorly in privacy, indicating that its token-masking objective does not sufficiently protect canary information in this setting.

Rewriting-based methods generally occupy intermediate regions of both panels, providing relatively balanced privacy--capability trade-offs. For both target models, privacy-aware rewriting achieves better privacy and capability ranks than standard rewriting. For rewriting, we use GPT-5.5 and Qwen3.5-4B as closed-source and open-weight rewrite models. Across both target models, GPT-5.5 achieves higher privacy ranks but lower capability ranks than Qwen3.5-4B, suggesting that its stronger privacy protection comes at some cost to capability.
However, rewriting-based methods require additional computation because each training record must pass through a generative model before finetuning. The resulting privacy and capability gains must be considered alongside this preprocessing cost. We further analyze rewriting in Section~\ref{sec:rewriting}.


\subsection{Canary Duplication}
\begin{figure}[t]
\centering
\includegraphics[width=1.0\linewidth]{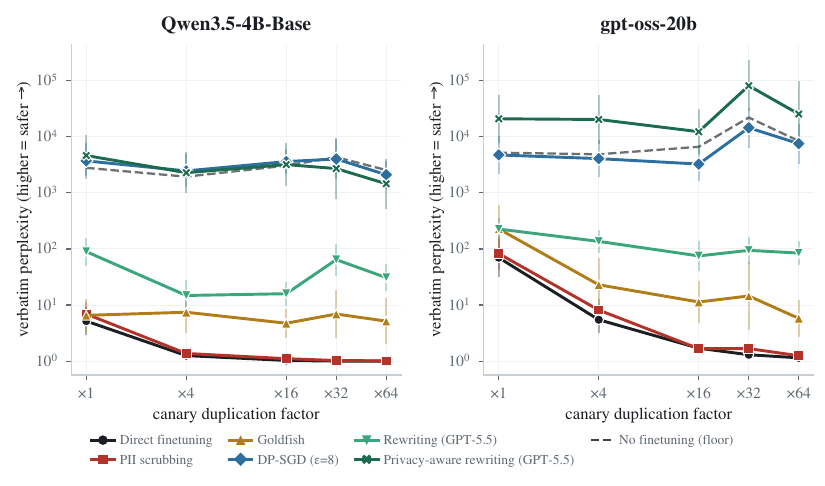}
\vspace{-6mm}
\caption{\textbf{Canary privacy leakage across levels of duplication.} 
Higher values indicate less privacy leakage, and error bars show 95\% confidence intervals over 20 canaries per group.}\
\label{fig:canary_duplication}
\vspace{-5mm}
\end{figure}

We examine how canary duplication affects privacy using five disjoint groups of 20 canaries, inserted at different frequencies ($\times1$, $\times4$, $\times16$, $\times32$, and $\times64$). Figure~\ref{fig:canary_duplication} and Figure~\ref{fig:canary_duplication_2} report verbatim and semantic perplexity across these duplication factors. 
Because each duplication level contains a different group of canaries, differences across levels can reflect both duplication and intrinsic canary difficulty. The no finetuning model provides a reference for this difficulty. For example, its higher perplexity at $\times32$ indicates that this group is intrinsically more difficult. We therefore interpret each trained condition relative to the no finetuning reference at the same duplication level.

Across both target models, PII scrubbing generally tracks direct finetuning and produces low perplexity at higher duplication levels, indicating that removing direct identifiers provides little protection against privacy leakage. In contrast, DP-SGD and GPT-5.5 privacy-aware rewriting maintain higher perplexity, even at high duplication levels. Goldfish is generally less sensitive to increasing duplication than direct finetuning. This relative stability is consistent with its deterministic, context-dependent masking. Because identical copies receive the same mask, duplicating a record does not add new training signals for tokens excluded from the loss.

\subsection{Rewriting}
\label{sec:rewriting}

\begin{figure}[t]
\centering
\includegraphics[width=1.0\linewidth]{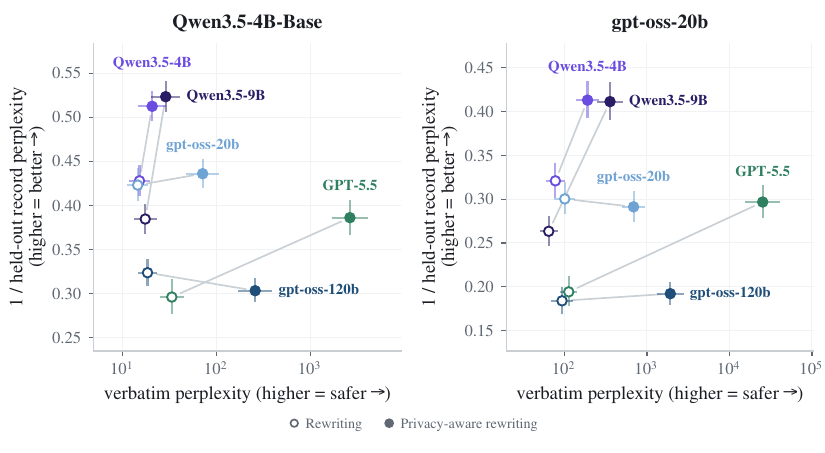}
\vspace{-5mm}
\caption{\textbf{Privacy--capability trade-offs across rewriters.} Open and filled markers denote rewriting and privacy-aware rewriting, with arrows connecting each pair. }\
\vspace{-7mm}
\label{fig:rewrite_combined}
\end{figure}

Previous results show that rewriting-based methods provide relatively balanced privacy--capability trade-offs. We therefore compare rewrite models from different model families, ablating the size, amount of test-time compute, and rewrite strategy to examine how much the exact rewriting details affect this trade-off.

\paragraph{Scaling Rewrite Model Capabilities improves Privacy}
As shown in Figure~\ref{fig:rewrite_combined}, adding the privacy-aware instruction improves the privacy rank for every rewrite model comparison shown in both panels. Capability improves in most cases. GPT-5.5 achieves the highest privacy rank in both target-model panels. Overall, larger rewrite models generally achieve higher privacy ranks than smaller rewrite models, particularly when given the privacy-aware instruction. This suggests that more capable rewrite models are better able to identify and replace distinctive details.

\begin{figure}[t]
\centering
\includegraphics[width=1.0\linewidth]{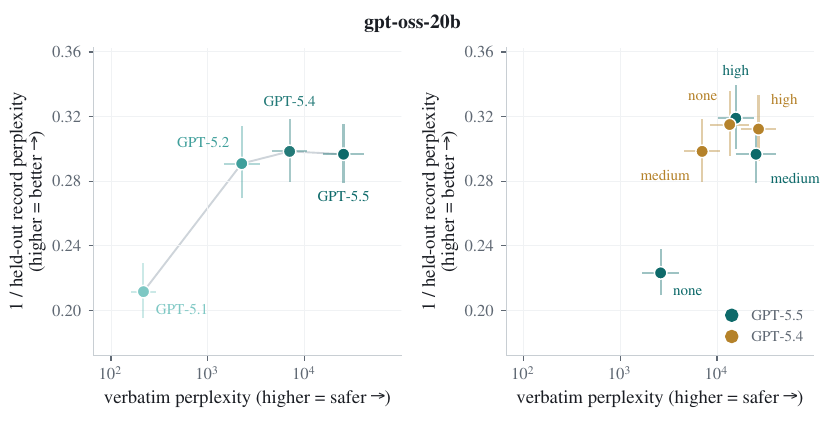}
\vspace{-5mm}
\caption{\textbf{Privacy--capability trade-offs for privacy-aware rewriting}
with gpt-oss-20b as the target model. 
The left shows different rewriting models, while the right shows different amounts of reasoning. Error bars denote 95\% confidence intervals.}\
\label{fig:oss_rewrite}
\vspace{-8mm}
\end{figure}

We further examine privacy-aware rewriting using several rewrite models, as shown in Figure~\ref{fig:oss_rewrite}. Across both target models, later GPT-5 variants consistently achieve better privacy protection, whereas capability does not improve monotonically. Capability generally increases from GPT-5.1 through GPT-5.4. These results suggest that using a stronger rewrite model can improve privacy protection without guaranteeing a corresponding improvement in capability.

%

\paragraph{Scaling Rewrite Test-time Compute Improves Privacy Tradeoffs}
We further vary the rewrite model's reasoning effort in the privacy-aware rewriting condition to examine how inference-time computation during rewriting affects privacy protection and capability. As shown in Figure~\ref{fig:oss_rewrite}, increasing reasoning effort does not monotonically improve either rank, and the best setting depends on the target model. Nevertheless, enabling reasoning generally provides a more favorable trade-off than the no-reasoning setting. These results suggest that some reasoning is beneficial, but increasing the reasoning budget does not consistently improve performance.


\paragraph{Summary-Based Rewriting}



\begin{figure}[t]
\centering
\includegraphics[width=1.0\linewidth]{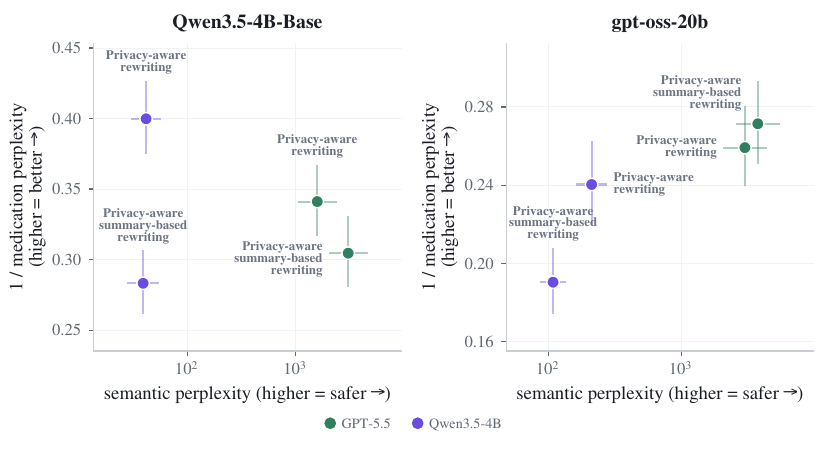}
\vspace{-9mm}
\caption{\textbf{Privacy--capability trade-offs for privacy-aware rewriting with and
without a summarization step.} 
Error bars denote 95\% confidence intervals.}\
\label{fig:summary_based_figure}
\vspace{-7mm}
\end{figure}

Summarization may improve privacy by discarding record-specific details while retaining useful clinical information. We therefore evaluate a rewriting technique that first summarizes each record and then expands the summary into a new record to control for formatting. We compare standard summarization with a privacy-aware variant that applies our uniqueness-aware instruction. Figure~\ref{fig:summary_based_figure} shows that, compared with privacy-aware rewriting alone, adding summarization generally yields similar or lower capability, potentially due to the loss of useful information. The effect on privacy leakage varies across models, suggesting no consistent improvement in the privacy--capability trade-off.

\begin{figure}[t]
\centering
\includegraphics[width=1.0\linewidth]{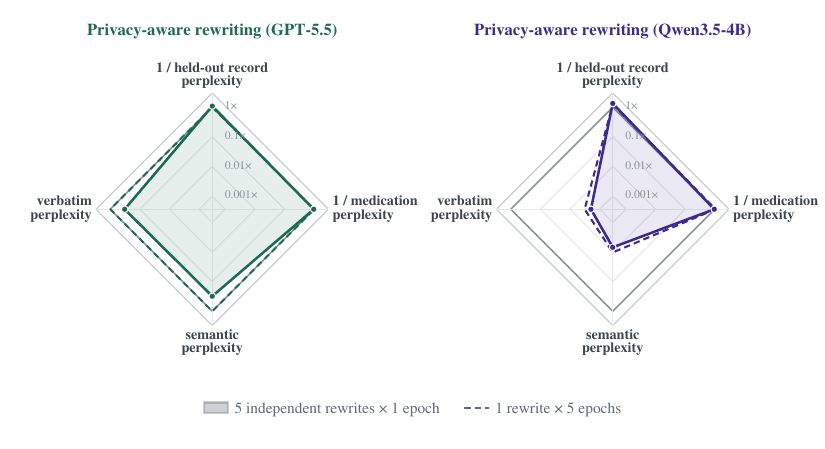}
\vspace{-9mm}
\caption{\textbf{Comparison of five independent rewrites trained for one epoch each and a
single rewrite trained for five epochs.} In both panels, the GPT-5.5
single-rewrite, five-epoch condition is normalized to \(1\times\); all other
values are reported as ratios to this reference. Higher values indicate
better performance on every axis.}
\vspace{-4mm}
\label{fig:independent_rewrite_vs_epochs}
\end{figure}

\paragraph{Independent Rewrites versus Repeated Training}
Training for multiple epochs repeatedly exposes the model to the same rewritten records. We therefore compare two settings: training on one rewrite for five epochs and training for one epoch on five independently generated rewrites. As shown in Figure~\ref{fig:independent_rewrite_vs_epochs}, the two radar profiles largely overlap, indicating similar performance across the evaluated metrics. The one-rewrite, five-epoch condition performs slightly better overall, suggesting that independent rewrites provide no clear advantage despite their additional generation cost.

\begin{figure}[h!]
\centering
\includegraphics[width=1.0\linewidth]{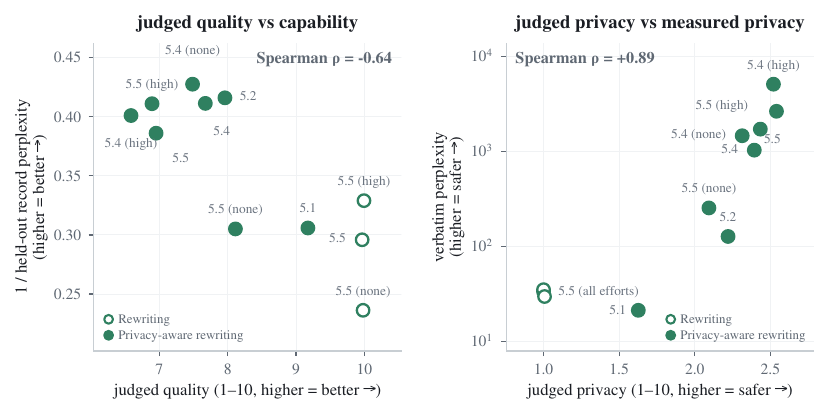}
\vspace{-5mm}
\caption{\textbf{Comparison of LLM-judged scores with measured outcomes using Qwen3.5-4B-Base as the target model.} Left: judged quality versus capability. Right: judged privacy versus verbatim perplexity. Point labels indicate the GPT-series rewriter and reasoning-effort setting.}\
\label{fig:llm_as_a_judge}
\vspace{-8mm}
\end{figure}

\subsection{Privacy and Capability Metrics for Data}
Our primary evaluation finetunes a target model on each rewritten corpus and measures privacy and capability. We also examine whether an LLM can predict these outcomes by inspecting the rewritten data alone. We use GPT-5.6-sol with xhigh reasoning to compare each original record with its corresponding rewrite and assign separate privacy and quality scores. To reduce confounding from cross-family stylistic differences, we restrict this analysis to GPT-series rewrite models.

As shown in Figure~\ref{fig:llm_as_a_judge}, judged privacy strongly correlates with verbatim perplexity (Spearman's $\rho=0.89$). Because higher verbatim perplexity indicates less privacy leakage, the judge successfully tracks the relative privacy ordering of the rewritten corpora. In contrast, judged quality correlates negatively with measured capability ($\rho=-0.64$), indicating that rewrites rated as higher quality do not necessarily produce more capable finetuned models. Thus, LLM judging may provide a useful secondary signal for privacy risk, but it cannot replace direct capability evaluation.
\section{Conclusion}
We presented a controlled evaluation of the privacy--capability trade-off
in language-model finetuning using a synthetic medical-record corpus. Among the evaluated methods, privacy-aware rewriting achieves the most balanced overall trade-off. Stronger rewrite models generally provided better privacy protection, although capability and the benefits of additional reasoning effort were not monotonic. We also find that LLM-judged privacy tracks verbatim leakage, whereas judged data quality does not reliably predict model capability. These findings establish privacy-aware rewriting as a promising direction for balancing privacy and capability.

\subsection*{AI use statement}

We used generative AI tools to help polish the writing and to draft parts of the text, such as figure captions. The research ideas in this work are our own and did not come
from AI tools. We reviewed and verified all AI-assisted text. The language models that serve as rewriters and as a judge in our experiments are part of the studied methods and
are described in Section~\ref{sec:methods}. We take full responsibility for the content
of this work, including any text, claims, or artifacts produced with the aid of
generative AI.

\subsection*{Reproducibility statement}
All data are fully synthetic: patient records are generated with Synthea from a fixed seed, and canaries are built by seeded code, so the full training corpus can be regenerated. Both fine-tuned models are open-weight (Qwen3.5-4B-Base and gpt-oss-20b), and every fine-tuned model uses the same LoRA setup (rank 16, $\alpha{=}32$, learning rate $2{\times}10^{-4}$, 2 epochs, 1,024-token sequences, seed 0); DP-SGD adds per-example clipping (norm 1.0) and Gaussian noise calibrated to $\varepsilon{=}8$, $\delta{=}10^{-5}$. API rewriters and the LLM judge are called with fixed model identifiers and reasoning-effort settings.  Prompts, hyperparameters, and metric definitions are given in the appendix.

\subsection*{Acknowledgements}
We are thankful to Om Thakkar and Christopher Ackerman for helpful discussion and comments on the paper. We also thank the MATS program for its support.

\bibliography{iclr2027_conference}
\bibliographystyle{iclr2027_conference}

\appendix

\clearpage

\section{Additional Results}
\subsection{privacy--Capability Trade-offs}
\begin{figure}[h]
\centering
\includegraphics[width=1.0\linewidth]{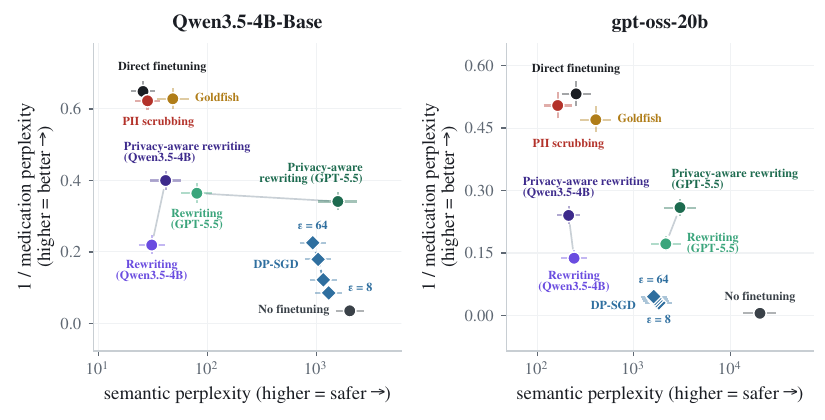}
\caption{\textbf{Privacy--capability trade-offs across methods.} Lines connect rewriting to privacy-aware rewriting for the same rewrite model. Error bars denote 95\% confidence intervals.}\
\label{fig:method_compare_details}
\end{figure}

\begin{figure}[h]
\centering
\includegraphics[width=1.0\linewidth]{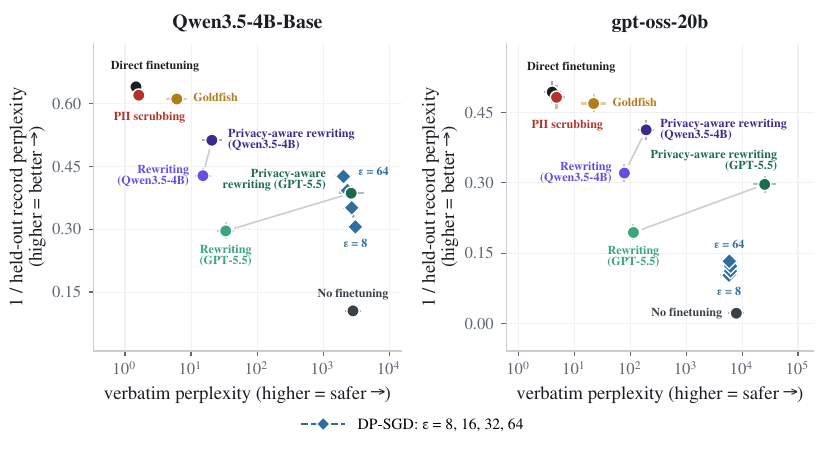}
\caption{\textbf{Privacy--capability trade-offs across methods.} Lines connect rewriting to privacy-aware rewriting for the same rewrite model. Error bars denote 95\% confidence intervals.}\
\label{fig:method_compare_details_2}
\end{figure}

Figures~\ref{fig:method_compare_details} and~\ref{fig:method_compare_details_2}
report the individual metric values for each method. Compared with direct finetuning, privacy-aware rewriting shows only small differences in the capability metrics but substantially higher verbatim and semantic perplexity, indicating reduced privacy leakage.

\begin{figure}[h]
\centering
\includegraphics[width=1.0\linewidth]{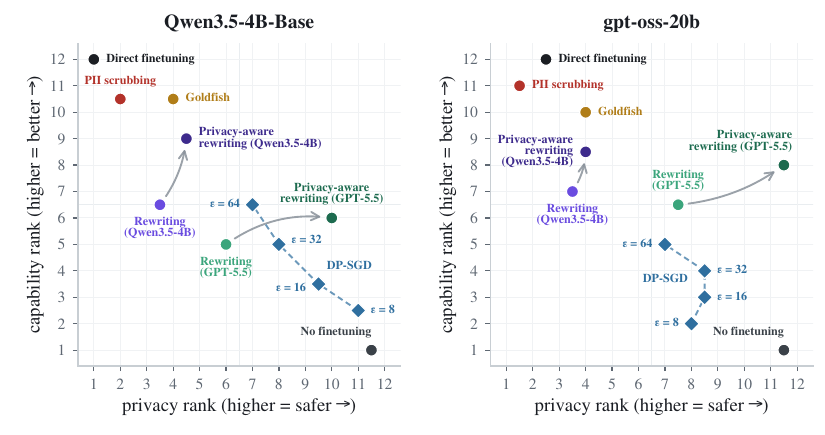}
\caption{\textbf{Privacy--capability trade-offs across methods.} Arrows connect rewriting to privacy-aware rewriting for the same rewrite model.}\
\label{fig:method_compare}
\end{figure}

\subsection{Canary Duplication}

\begin{figure}[h]
\centering
\includegraphics[width=1.0\linewidth]{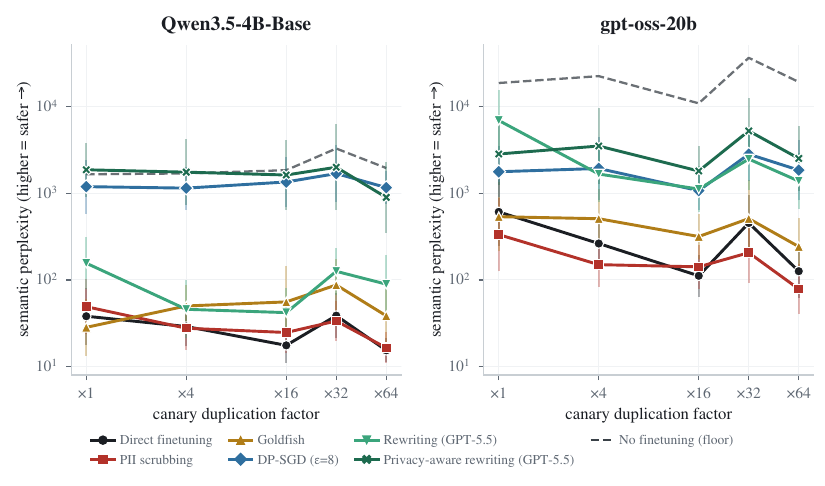}
\caption{\textbf{Canary privacy leakage across duplication factors.} Rows show Qwen3.5-4B-Base and gpt-oss-20b; columns show semantic perplexity. Higher values indicate less privacy leakage, and error bars show 95\% confidence intervals over 20 canaries per group.}\
\label{fig:canary_duplication_2}
\end{figure}

\subsection{Scaling Rewrite Model Capabilities}

\begin{figure}[h]
\centering
\includegraphics[width=1.0\linewidth]{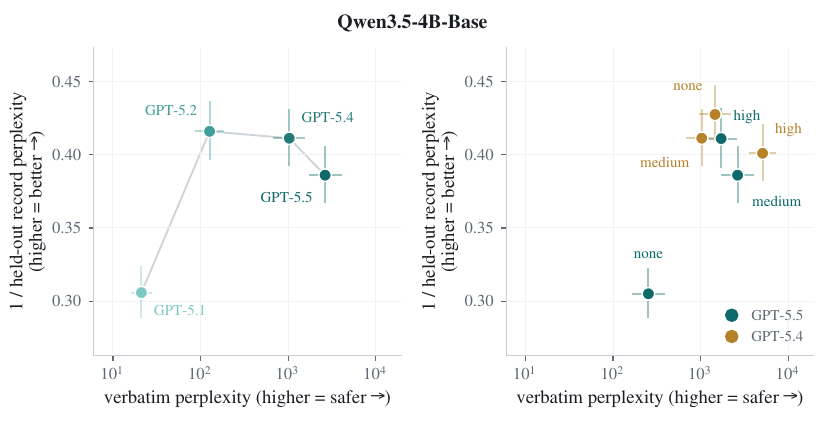}
\caption{\textbf{Privacy--capability trade-offs for privacy-aware rewriting}
with Qwen3.5-4B-Base as the target model. 
The left shows different rewriting models, while the right shows different amounts of reasoning. Error bars denote 95\% confidence intervals.}\
\label{fig:qwen_rewrite}
\end{figure}

\begin{figure}[h]
\centering
\includegraphics[width=1.0\linewidth]{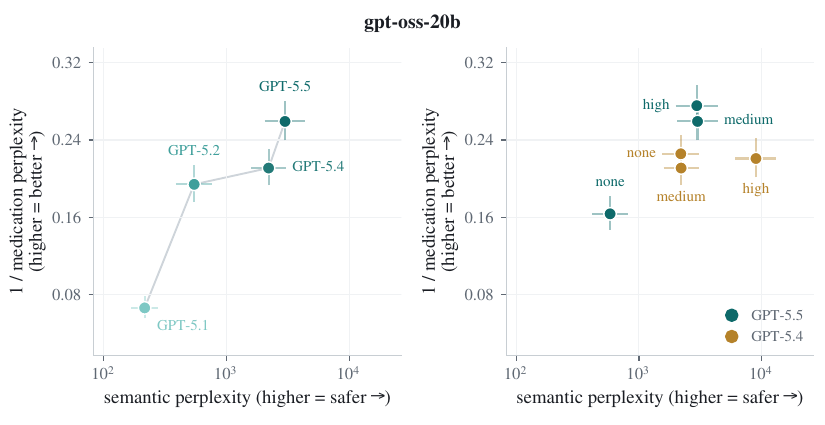}
\caption{\textbf{Privacy--capability trade-offs for privacy-aware rewriting}
with gpt-oss-20b as the target model. 
The left shows different rewriting models, while the right shows different amounts of reasoning. Error bars denote 95\% confidence intervals.}\
\label{fig:gpt_rewrite_2}
\end{figure}

\begin{figure}[h]
\centering
\includegraphics[width=1.0\linewidth]{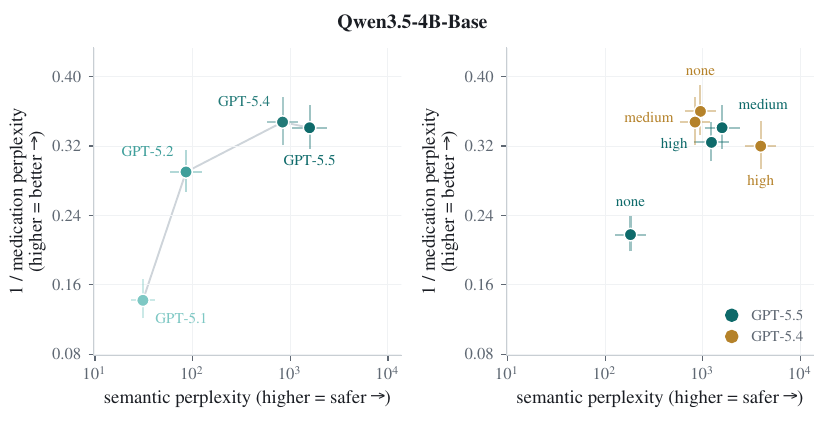}
\caption{\textbf{Privacy--capability trade-offs for privacy-aware rewriting}
with Qwen3.5-4B-Base as the target model. 
The left shows different rewriting models, while the right shows different amounts of reasoning. Error bars denote 95\% confidence intervals.}\
\label{fig:qwen_rewrite_2}
\end{figure}

Figure~\ref{fig:qwen_rewrite} shows that, with Qwen3.5-4B as the target model,
stronger rewrite models generally improve privacy protection but do not
consistently improve the target model's capability.
Figures~\ref{fig:gpt_rewrite_2} and~\ref{fig:qwen_rewrite_2} show similar
trends when using semantic perplexity and inverse medication perplexity
as the privacy and capability metrics.

\subsection{Scaling Rewrite Test-time Compute}
Figure~\ref{fig:qwen_rewrite} shows that, with Qwen3.5-4B as the target model,
increasing reasoning effort does not consistently improve privacy or
capability, although reasoning-enabled settings generally provide better
privacy protection than the none setting.
Figures~\ref{fig:gpt_rewrite_2} and~\ref{fig:qwen_rewrite_2} show similar
trends when using semantic perplexity and inverse medication perplexity
as the privacy and capability metrics.

\subsection{Summary-Based Rewriting}
\begin{figure}[t]
\centering
\includegraphics[width=1.0\linewidth]{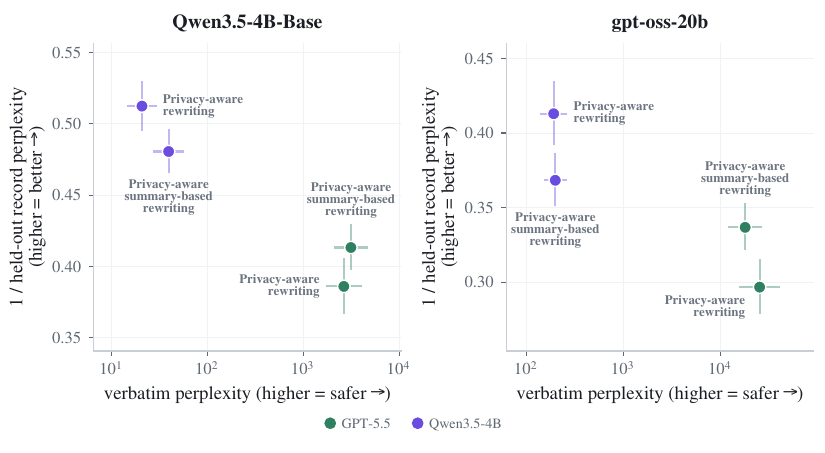}
\vspace{-9mm}
\caption{\textbf{Privacy--capability trade-offs for privacy-aware rewriting with and
without a summarization step.} 
Error bars denote 95\% confidence intervals.}\
\label{fig:summary_based_figure_2}
\vspace{-7mm}
\end{figure}

As shown in Figure~\ref{fig:summary_based_figure_2}, the effects of summarization remain model-dependent. Held-out record perplexity may partly reflect the formatting guidance provided during expansion, since the
auxiliary examples share a similar structure with the held-out records. The effect on verbatim perplexity also varies across target models. Overall, privacy-aware summary-based rewriting remains a practical alternative to privacy-aware rewriting without summarization
when the finetuning compute budget is limited.

\subsection{Independent Rewrites versus Repeated Training}
\begin{figure}[t]
\centering
\includegraphics[width=1.0\linewidth]{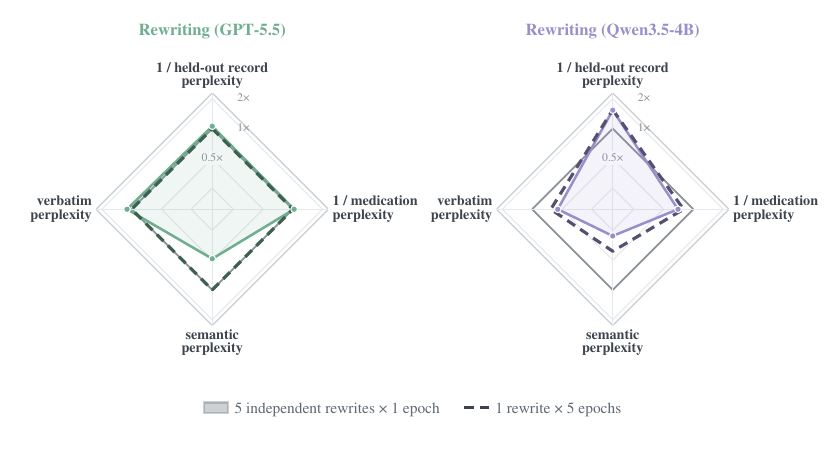}
\vspace{-9mm}
\caption{\textbf{Comparison of five independent rewrites trained for one epoch each and a
single rewrite trained for five epochs.} In both panels, the GPT-5.5
single-rewrite, five-epoch condition is normalized to \(1\times\); all other
values are reported as ratios to this reference. Higher values indicate
better performance on every axis.}
\vspace{-3mm}
\label{fig:independent_rewrite_vs_epochs_2}
\end{figure}

Figure~\ref{fig:independent_rewrite_vs_epochs_2} shows that, for rewriting, training on one rewrite per record for five epochs outperforms training on
five independent rewrites per record for one epoch.

\subsection{Privacy and capability metrics for data}
\begin{figure}[h!]
\centering
\includegraphics[width=1.0\linewidth]{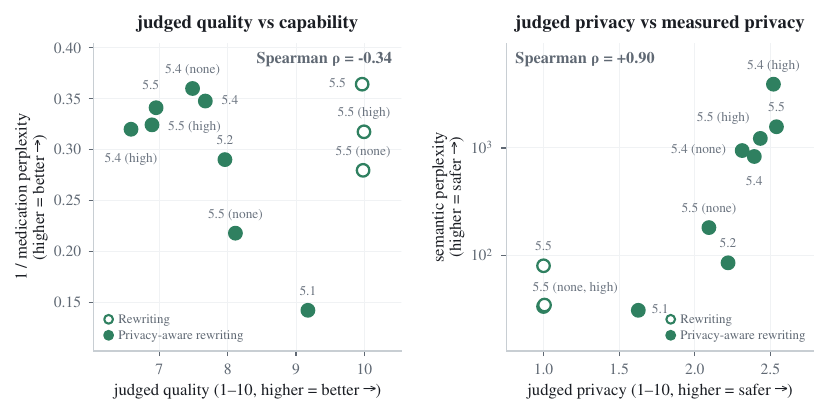}
\caption{\textbf{Comparison of LLM-judged scores with measured outcomes using Qwen3.5-4B-Base as the target model.} Left: judged quality versus capability. Right: judged privacy versus verbatim perplexity. Point labels indicate the GPT-series rewriter and reasoning-effort setting.}\
\label{fig:llm_as_a_judge_2}
\vspace{-8mm}
\end{figure}

\begin{figure}[h!]
\centering
\includegraphics[width=1.0\linewidth]{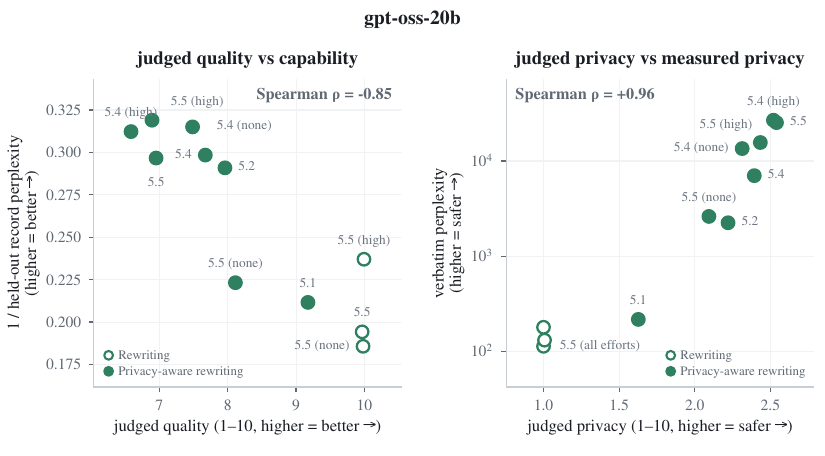}
\caption{\textbf{Comparison of LLM-judged scores with measured outcomes using gpt-oss-20b as the target model.} Left: judged quality versus capability. Right: judged privacy versus verbatim perplexity. Point labels indicate the GPT-series rewriter and reasoning-effort setting.}\
\label{fig:llm_as_a_judge_oss}
\vspace{-8mm}
\end{figure}

\begin{figure}[h!]
\centering
\includegraphics[width=1.0\linewidth]{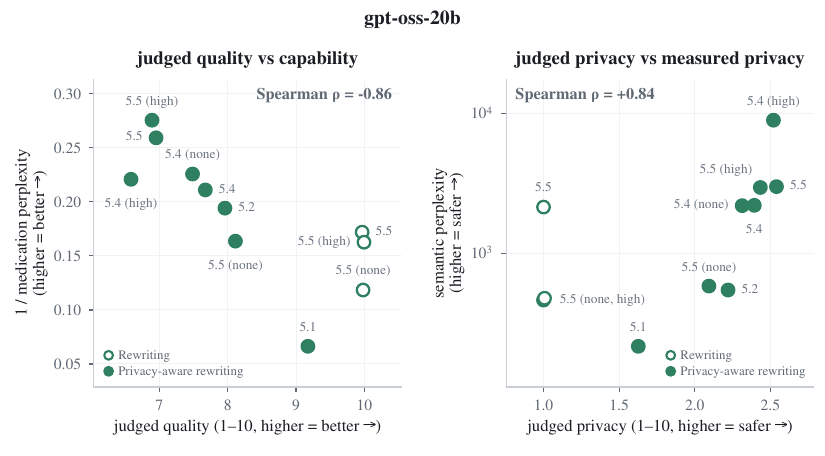}
\caption{\textbf{Comparison of LLM-judged scores with measured outcomes using gpt-oss-20b as the target model.} Left: judged quality versus capability. Right: judged privacy versus verbatim perplexity. Point labels indicate the GPT-series rewriter and reasoning-effort setting.}\
\label{fig:llm_as_a_judge_oss_2}
\vspace{-6mm}
\end{figure}

Figure~\ref{fig:llm_as_a_judge_2} compares judged quality with medication perplexity and judged privacy with semantic perplexity. The results support
the same conclusion: LLM-judged privacy tracks measured privacy, whereas judged quality does not reliably predict model capability. Figures~\ref{fig:llm_as_a_judge_oss} and~\ref{fig:llm_as_a_judge_oss_2}
show similar trends with gpt-oss-20b as the target model.

\subsection{Correlation of Privacy Metrics}

\begin{figure}[h!]
\centering
\includegraphics[width=1.0\linewidth]{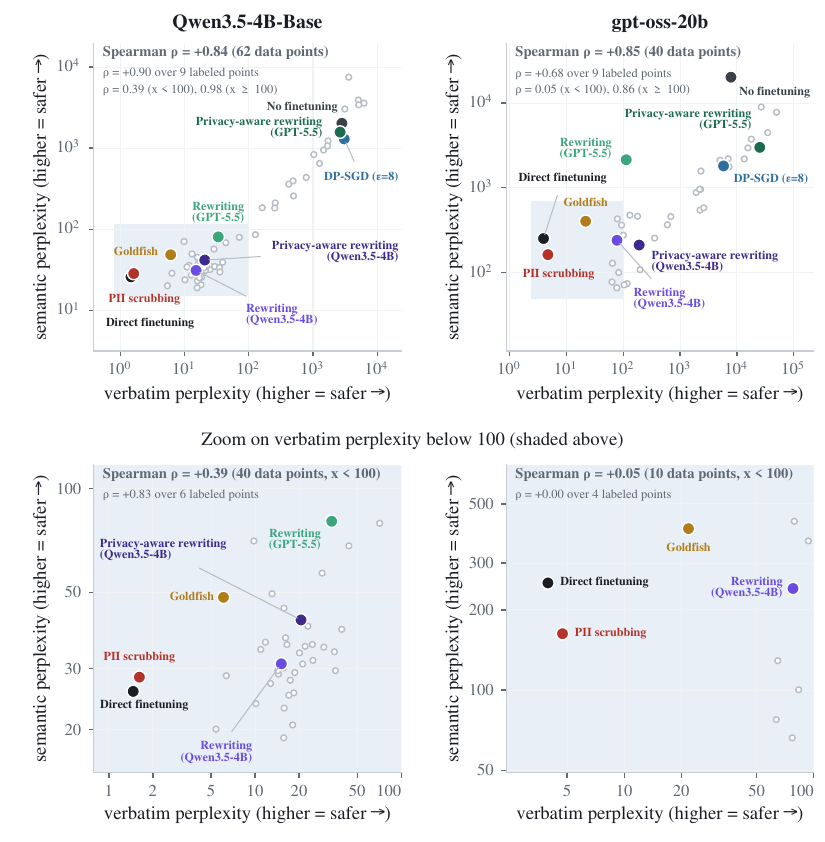}
\vspace{-4mm}
\caption{\textbf{The correlation between two privacy metrics.} Verbatim versus semantic
perplexity of the planted secret for each method configuration. Labeled points are the nine methods; gray points are rewriting and summary
variants. The second row zooms into the shaded region below verbatim perplexity 100,
where agreement drops to $\rho = 0.39$ and $0.05$.}
\label{fig:corr_privacy}
\vspace{-2mm}
\end{figure}

We examine the correlation between verbatim and semantic perplexity as shown in Figure~\ref{fig:corr_privacy}, grouping results by verbatim perplexity. For Qwen3.5-4B-Base, the correlation is 0.84, but the correlation is 0.39 below a verbatim perplexity of 100 and 0.98 above this threshold. For gpt-oss-20b, the correlation is 0.85, but the correlation is 0.05 below a verbatim perplexity of 100 and 0.68 above this threshold. These results support reporting both metrics, as they provide complementary measurements of privacy leakage, particularly when verbatim perplexity is low.

\subsection{Correlation of Capability Metrics}
\begin{figure}[h!]
\centering
\includegraphics[width=1.0\linewidth]{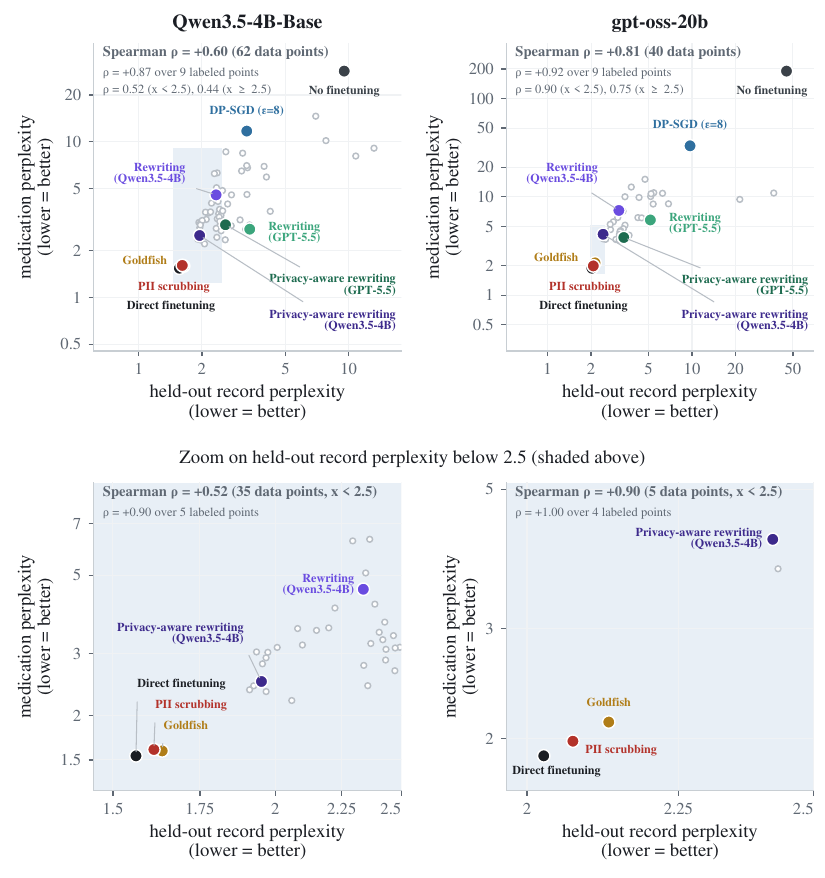}
\caption{\textbf{The correlation between two capability metrics.} Held-out record
perplexity versus medication perplexity for each method configuration. Labeled points are the nine methods; gray points are rewriting and
summary variants. The second row zooms into the shaded region below held-out
perplexity 2.5, where agreement is similar ($\rho = 0.52$ and $0.90$, over 35 and 5 points).}
\label{fig:corr_capability}
\vspace{-2mm}
\end{figure}

We also examine the correlation between held-out record perplexity and medication perplexity as shown in Figure~\ref{fig:corr_capability}. For Qwen3.5-4B-Base, the correlation is 0.60, but the correlation is 0.52 when held-out record perplexity is below 2.5 and 0.44 above this threshold.
For gpt-oss-20b, the correlation is 0.81, but the correlation is 0.90 when held-out record perplexity is below 2.5 and 0.75 above this threshold. These results support reporting both metrics, as they capture different aspects of model capability.

\subsection{Probability of generating the coined medication}
Perplexity is the inverse of the model's average per-token probability, which makes differences between methods hard to read. We therefore also report the probability that a model generates the coined medication given the record up to it: the product of the model's probabilities for the medication's tokens. This is the chance that a single sampled continuation reproduces the coined medication.

\begin{figure}[h!]
\centering
\includegraphics[width=1.0\linewidth]{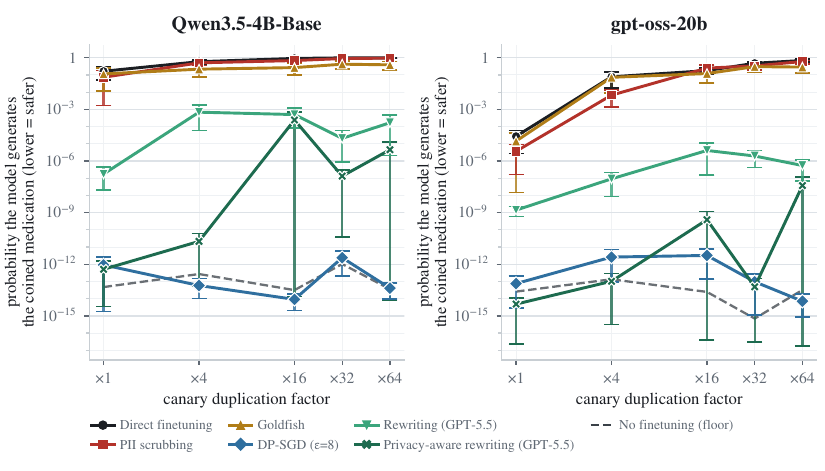}
\caption{\textbf{Probability of generating the coined medication.} Probability that each target model generates the coined medication given the record up to it, against how many times the canary appeared in training. Each point is the mean over the 20 canaries at that duplication level, with 95\% confidence intervals.}
\label{fig:probability}
\vspace{-2mm}
\end{figure}

Figure~\ref{fig:probability} shows this probability against the number of times each canary appeared in training. After direct finetuning on Qwen3.5-4B-Base, a coined medication seen once is generated 16\% of the time and one seen 64 times 99\% of the time; PII scrubbing follows closely (7\% to 94\%), and Goldfish levels off at 30--40\%. gpt-oss-20b needs duplication before it memorizes: after direct finetuning, a coined medication seen once is generated about once in 37{,}000 samples, rising to 69\% at $\times$64. Rewriting with GPT-5.5 lowers the probability to between $10^{-9}$ and $10^{-3}$, and DP-SGD stays on the no finetuning floor, below $10^{-11}$, at every duplication level. Under privacy-aware rewriting with GPT-5.5 the typical canary also sits at the floor; the jumps at $\times$16 and $\times$64 each come from a single leaked canary (on Qwen3.5-4B-Base at $\times$16, one canary is generated 0.5\% of the time and the next only $10^{-11}$ of the time).

\subsection{Likelihood-based exposure of canary medications}

To assess how strongly a model favors a planted medication over alternative completions, we complement our perplexity measurements with the exposure metric introduced by \cite{carlini2019secret}. For each canary, we condition on the original record prefix ending immediately before its medication name and score candidate names using summed token log probabilities. The candidate set contains the planted name and 6,300 unplanted alternatives drawn from the same coined-name space, excluding the other 99 planted medications. If the true name ranks \(r\) among these 6,301 candidates, its exposure is
\[
E=\log_2(6{,}301)-\log_2 r.
\]
We report mean exposure across the 100 canaries and the proportion whose planted name ranks first.

Table~\ref{tab:exposure} shows exposure results on both target models. Mean exposure of direct finetuning 
reaches 12.22 bits for Qwen3.5-4B-Base and 10.98 bits for gpt-oss-20b, with the planted medication ranked first for 86\% and 73\% of canaries. PII scrubbing produces similar exposure, indicating limited protection of the medication content. Goldfish reduces exposure relative to direct finetuning, but the planted medication remains the highest-ranked candidate for 53\% and 37\% of canaries.

Rewriting reduces exposure on both target models, and privacy-aware instructions further reduce it for both rewriters. The reduction is larger with GPT-5.5 than with Qwen3.5-4B. Privacy-aware GPT-5.5 rewriting yields mean exposures of 2.21 and 1.71 bits. DP-SGD achieves the lowest mean exposure among the trained conditions—1.74 and 1.56 bits—with no rank-1 canaries observed. Its averages are close to the corresponding no finetuning values of 1.49 and 1.39 bits.

\begin{table}[t]
\centering
\small
\caption{\textbf{Secret-Sharer exposure of coined medication names.} Each planted coined medication is ranked against 6,300 unplanted alternatives by its whole-name conditional likelihood given the original record prefix ending immediately before the medication. We report mean exposure, \(\log_2(6{,}301/\mathrm{rank})\), across 100 canaries and the percentage ranked first within their candidate sets (“Rank 1”). Exposure ranges from 0 to 12.62 bits; uniformly random ranks have an expected mean exposure of approximately 1.44 bits. Lower values indicate less exposure.}
\label{tab:exposure}
\begin{tabular}{lcccc}
\toprule
 & \multicolumn{2}{c}{Qwen3.5-4B-Base} & \multicolumn{2}{c}{gpt-oss-20b} \\
\cmidrule(lr){2-3}\cmidrule(lr){4-5}
Method & Exposure & Rank 1 & Exposure & Rank 1 \\
\midrule
No finetuning & 1.49 & 0\% & 1.39 & 0\% \\
Direct fine-tuning & 12.22 & 86\% & 10.98 & 73\% \\
Goldfish ($k{=}4$) & 10.38 & 53\% & 8.90 & 37\% \\
DP-SGD ($\varepsilon{=}8$) & 1.74 & 0\% & 1.56 & 0\% \\
PII scrubbing & 12.16 & 87\% & 10.76 & 68\% \\
Rewriting (GPT-5.5) & 6.86 & 12\% & 4.46 & 4\% \\
Privacy-aware rewriting (GPT-5.5) & 2.21 & 5\% & 1.71 & 1\% \\
Rewriting (Qwen3.5-4B) & 7.85 & 30\% & 4.91 & 7\% \\
Privacy-aware rewriting (Qwen3.5-4B) & 6.94 & 24\% & 4.03 & 6\% \\
\bottomrule
\end{tabular}
\end{table}

\section{Additional Experiment Details}

\subsection{Summary-based Rewriting}
The prompts for summary-based rewriting closely follow those used for rewriting and are provided in Table~\ref{tab:prompts_summary}.

\begin{table*}[t]
\caption{Generation prompts. The two summary-based methods summarize first and then expand the summary back into a full record with the expansion prompt,
which additionally receives two format exemplars drawn from a fresh Synthea
seed, never from the private cohort. The privacy-aware summary prompt is the generic summary instruction followed by the privacy-aware rewriting
instruction.}
\centering
\small
\renewcommand{\arraystretch}{1.2}
\begin{tabular}{@{}p{0.18\textwidth}p{0.77\textwidth}@{}}
\toprule
\textbf{Method} & \textbf{Prompt} \\
\midrule
\textbf{Summary-based rewriting}
& \itshape Summarize the following clinical record in a few sentences,
keeping the clinically essential information (active diagnoses, key
treatments, current medications): \\

\addlinespace[3pt]
\textbf{Privacy-aware summary-based rewriting}
& \itshape Summarize the following clinical record in a few sentences,
keeping the clinically essential information. Replace any detail that is
rare, distinctive, or unusual enough that it could help identify this
specific patient with a different but plausible fictional value; leave
common, generic clinical information unchanged. Use realistic surrogate
values, not placeholders. \\

\midrule
\textbf{Expansion prompt}
& \itshape Below is a clinical summary of a patient. Reconstruct a complete,
realistic patient record consistent with it, in exactly the format of the
example records shown. Include every diagnosis and medication the summary
mentions; invent realistic demographics and identifiers (name, address, SSN,
DOB consistent with the age) --- the summary does not contain them; add
plausible routine detail where the summary is silent. \\
\bottomrule
\end{tabular}
\label{tab:prompts_summary}
\end{table*}

\section{Case study}
\label{sec:case-study}
We examine how different methods process an individual canary and how the resulting training data affect privacy leakage. Canary-0091 describes the fabricated patient \hlid{Lenora325 Daugherty83} and contains the coined medication \hlsecret{Velfazvastatin}. It appears 64 times in the training corpus. We select this representative canary whose perplexities are closest to the median results across metrics on Qwen3.5-4B-Base.
Table~\ref{tab:case-trace} presents the transformed records, medication and name retention counts, and privacy leakage results.

Unless stated otherwise, the values below refer to Qwen3.5-4B-Base, with verbatim and
semantic perplexity reported in that order.

\providecommand{\casecjk}[2]{{\rmfamily\itshape[Chinese: #2]}}
{\setlength{\tabcolsep}{3pt}\setlength{\LTcapwidth}{\linewidth}

\clearpage
\begin{longtable}{@{}>{\scriptsize\raggedright\arraybackslash}p{1.02in}>{\scriptsize\raggedright\arraybackslash\frenchspacing\hyphenpenalty=10000 }p{\dimexpr\linewidth-1.02in-2\tabcolsep\relax}@{}}
\caption{Canary-0091 across methods. We show the full text of one processed copy of the record, as it reaches training. Direct fine-tuning, Goldfish and DP-SGD train on the same record, shown once; shaded tokens are excluded from the loss by Goldfish only, under the Qwen3.5 tokenizer. PII scrubbing changes only the header, so only the header is shown. The coined medication is highlighted in yellow, original identity spans in blue, replacement identities in blue italics, and scrubbing placeholders in red. Rewriting and summary methods process each of the 64 copies independently; we show the first draw whose handling of the coined medication, kept or removed, matches the majority of the 64 draws.}
\label{tab:case-trace}\\
\toprule
Method & What reaches training (full text)\\
\midrule
\endfirsthead
\caption[]{(continued)}\\
\toprule
Method & What reaches training (full text)\\
\midrule
\endhead
\bottomrule
\endfoot
\bottomrule
\endlastfoot
\multicolumn{2}{@{}l}{\scriptsize\itshape Training-stage methods: the record itself reaches the fit}\\*[1pt]
Direct fine-tuning,\newline Goldfish ($k{=}4$),\newline DP-SGD ($\varepsilon{=}8$) & {Patient \hlid{Lenora325 Daugherty83}, 29yo F, white/nonhispanic\hlmask{.} Address: \hlidlong{1\hlmask{0}38 Quigley Mall Suite 2\hlmask{6}\hlmask{,} Medford, Massachusetts \hlmask{0}2\hlmask{1}5\hlmask{5}}. SSN: \hlid{9\hlmask{9}9\hlmask{-}\hlmask{3}4-\hlmask{2}\hlmask{9}53}, DOB \hlid{1996-08-1\hlmask{8}}. Active problems:\hlmask{ Housing} unsatisf\hlmask{actory} (finding\hlmask{);} Received higher education\hlmask{ (}finding); Pred\hlmask{i}abetes (finding); Anemia (disorder); Medication review due\hlmask{ (}situation); Part-time employment (finding\hlmask{);}\hlmask{ Stress}\hlmask{ (}finding); Normal pregnancy (finding\hlmask{);} Misc\hlmask{ar}\hlmask{riage} in\hlmask{ first} trimester (disorder);\hlmask{ Bl}ighted\hlmask{ ov}um\hlmask{ (}disorder); Past\hlmask{ pregnancy} history of miscar\hlmask{riage} (situation); Body mass index \hlmask{3}0+ - obesity (finding\hlmask{);}\hlmask{ Full}\hlmask{-time} employment\hlmask{ (}finding);\hlmask{ Ging}ivitis (\hlmask{dis}order); Acute\hlmask{ bron}chitis (dis\hlmask{order}); Social isolation\hlmask{ (}\hlmask{finding}); Suspected\hlmask{ disease} caused\hlmask{ by} Severe\hlmask{ acute} respiratory coronavirus \hlmask{2} (situation); Sput\hlmask{um}\hlmask{ finding} (finding\hlmask{);} Fever (\hlmask{finding}\hlmask{);} Disease\hlmask{ caused}\hlmask{ by} severe acute\hlmask{ respiratory}\hlmask{ syndrome}\hlmask{ coronavirus} \hlmask{2} (dis\hlmask{order}); Pne\hlmask{um}onia (\hlmask{dis}order\hlmask{);} Hypoxemia (disorder\hlmask{);} Respir\hlmask{atory} distress\hlmask{ (}finding); Ischemic heart disease (disorder); Abnormal findings diagnostic imaging heart\hlmask{+}\hlmask{cor}onary\hlmask{ circul}at\hlmask{ (}finding); History of coronary artery bypass\hlmask{ graft}ing (situation); Hyperlip\hlmask{id}emia (disorder); Has a criminal\hlmask{ record}\hlmask{ (}finding); Gingival disease (disorder). Medications: sodium\hlmask{ fluoride} 0.0272 MG\hlmask{/M}G Oral Gel\hlmask{;} \{2\hlmask{8} (norethindrone\hlmask{ }0\hlmask{.}35 MG Oral\hlmask{ Tablet}) \}\hlmask{ Pack} [J\hlmask{oliv}ette 28 Day]; Acetaminophen\hlmask{ }2\hlmask{1}.7 MG\hlmask{/}ML /\hlmask{ D}ext\hlmask{rom}\hlmask{eth}orphan Hydrobromide 1 MG\hlmask{/}ML\hlmask{ /} do\hlmask{x}ylamine\hlmask{ succ}inate \hlmask{0}.4\hlmask{1}7 MG/\hlmask{ML} Oral\hlmask{ Solution}\hlmask{;} \{7 (ethinyl estradi\hlmask{ol} 0.01 MG\hlmask{ Oral} Tablet)\hlmask{ /}\hlmask{ }84 (ethinyl estr\hlmask{adi}ol\hlmask{ }\hlmask{0}.0\hlmask{3} MG\hlmask{ /} lev\hlmask{on}orgestrel \hlmask{0}.1\hlmask{5} MG Oral Tablet\hlmask{)} \}\hlmask{ Pack}\hlmask{ [}Season\hlmask{ique}]; 0\hlmask{.}4\hlmask{ ML} Enoxaparin sodium\hlmask{ }100\hlmask{ MG}/\hlmask{ML} Prefilled\hlmask{ Sy}\hlmask{ringe}; Ac\hlmask{et}\hlmask{amin}ophen\hlmask{ }50\hlmask{0}\hlmask{ MG}\hlmask{ Oral} Tablet;\hlmask{ lev}onorgestrel 0\hlmask{.}00\hlmask{0}81\hlmask{3} MG/HR\hlmask{ In}\hlmask{tra}\hlmask{uter}ine System [Lilet\hlmask{ta}]; Clopidogrel \hlmask{7}\hlmask{5} MG Oral\hlmask{ Tablet}\hlmask{;} Simvastatin 20\hlmask{ MG} Oral Tablet; \hlmask{2}\hlmask{4}\hlmask{ HR} metoprol\hlmask{ol}\hlmask{ succ}inate 100 MG\hlmask{ Extended}\hlmask{ Release} Oral Tablet; Nitroglycer\hlmask{in}\hlmask{ }0\hlmask{.}4 MG\hlmask{/}ACTUAT M\hlmask{uc}osal Spray; c\hlmask{ef}az\hlmask{olin} \hlmask{2}000 MG\hlmask{ Injection};\hlmask{ }1 ML\hlmask{ he}parin sodium \hlmask{ porc}ine \hlmask{5}000\hlmask{ UNT}/ML Injection; 25\hlmask{ ML} protamine sulfate (US\hlmask{P}) 1\hlmask{0}\hlmask{ MG}/ML Injection; 2 ML\hlmask{ Ond}ans\hlmask{et}ron \hlmask{2} MG/\hlmask{ML} Injection; Mid\hlmask{azol}am 1\hlmask{ MG}\hlmask{/}ML Injectable\hlmask{ Solution}; \hlmask{1}\hlmask{0}0 ML Prop\hlmask{of}\hlmask{ol}\hlmask{ }10 MG/\hlmask{ML} Injection; Rocuronium bromide \hlmask{1}\hlmask{0} MG\hlmask{/}ML Injectable\hlmask{ Solution}; isof\hlmask{l}\hlmask{ur}ane\hlmask{ }\hlmask{9}\hlmask{9}\hlmask{9}\hlmask{ MG}\hlmask{/}ML Inhalation Solution\hlmask{;} 10 ML Alfentanil \hlmask{0}.5 MG/ML Injection;\hlmask{ lev}onorgest\hlmask{rel} \hlmask{0}\hlmask{.}000\hlmask{7}29 MG\hlmask{/}HR Intrauter\hlmask{ine} System [Kyle\hlmask{ena}\hlmask{];} \{2\hlmask{1}\hlmask{ (}eth\hlmask{in}yl estradi\hlmask{ol} \hlmask{0}.03\hlmask{ MG}\hlmask{ /} lev\hlmask{on}orgestrel \hlmask{0}.1\hlmask{5} MG Oral Tablet\hlmask{)} / 7 (inert ingredients 1\hlmask{ MG}\hlmask{ Oral}\hlmask{ Tablet}) \}\hlmask{ Pack} [Le\hlmask{v}ora \hlmask{0}\hlmask{.}1\hlmask{5}/3\hlmask{0} \hlmask{2}8 Day]; Simvast\hlmask{atin}\hlmask{ }10 MG Oral Tablet; \{24 (\hlmask{dro}sp\hlmask{ire}none \hlmask{3} MG\hlmask{ /} ethinyl estradi\hlmask{ol} 0\hlmask{.}02 MG Oral Tablet) / 4 (inert ingredients\hlmask{ }1\hlmask{ MG} Oral\hlmask{ Tablet}) \} Pack [Y\hlmask{az} 28 Day]; \hlsecret{Velfazvastatin} 2\hlmask{5}0 MG Injectable Solution.}\\ \addlinespace[4pt]
PII scrubbing & {Patient\allowbreak\hltag{[PRIVATE\_PERSON]}\allowbreak\hltag{[PRIVATE\_PERSON]}, 29yo F, white/nonhispanic. Address: \hltag{[PRIVATE\_ADDRESS]}\allowbreak\hltag{[PRIVATE\_ADDRESS]}. SSN: \hltag{[ACCOUNT\_NUMBER]}\allowbreak\hltag{[ACCOUNT\_NUMBER]}, DOB \hlid{1996-08-18}.} \textit{The rest of the record is unchanged.}\\ \addlinespace[4pt]
\addlinespace[6pt]\multicolumn{2}{@{}l}{\scriptsize\itshape Data-stage rewriters: one rewrite per copy reaches the fit (one of 64 draws shown)}\\*[1pt]
GPT-5.5 rewriting & {Clinical record summary for \hlid{Lenora325 Daugherty83}: 29-year-old white, non-Hispanic female. Date of birth: \hlid{1996-08-18}. SSN: \hlid{999-34-2953}. Residence: \hlidlong{1038 Quigley Mall Suite 26, Medford, MA 02155}.}\\
 & {Active conditions/problems documented include: unsatisfactory housing; higher education received; prediabetes; anemia; medication review due; part-time employment; stress; normal pregnancy; first-trimester miscarriage; blighted ovum; history of miscarriage in a prior pregnancy; BMI $\geq$30 consistent with obesity; full-time employment; gingivitis; acute bronchitis; social isolation; suspected illness due to Severe acute respiratory coronavirus 2; sputum-related finding; fever; confirmed disease caused by severe acute respiratory syndrome coronavirus 2; pneumonia; hypoxemia; respiratory distress; ischemic heart disease; abnormal diagnostic imaging findings involving the heart and coronary circulation; history of coronary artery bypass grafting; hyperlipidemia; criminal record; and gingival disease.}\\
 & {Current or recorded medications include: sodium fluoride 0.0272 mg/mg oral gel; Jolivette 28 Day pack containing 28 norethindrone 0.35 mg oral tablets; oral solution containing acetaminophen 21.7 mg/mL, dextromethorphan hydrobromide 1 mg/mL, and doxylamine succinate 0.417 mg/mL; Seasonique pack containing 7 ethinyl estradiol 0.01 mg oral tablets and 84 ethinyl estradiol 0.03 mg/levonorgestrel 0.15 mg oral tablets; 0.4 mL enoxaparin sodium 100 mg/mL prefilled syringe; acetaminophen 500 mg oral tablet; Liletta levonorgestrel 0.000813 mg/hr intrauterine system; clopidogrel 75 mg oral tablet; simvastatin 20 mg oral tablet; 24-hour metoprolol succinate 100 mg extended-release oral tablet; nitroglycerin 0.4 mg/actuation mucosal spray; cefazolin 2000 mg injection; 1 mL heparin sodium porcine 5000 units/mL injection; 25 mL protamine sulfate USP 10 mg/mL injection; 2 mL ondansetron 2 mg/mL injection; midazolam 1 mg/mL injectable solution; propofol 10 mg/mL injection in a 100 mL quantity; rocuronium bromide 10 mg/mL injectable solution; isoflurane 999 mg/mL inhalation solution; 10 mL alfentanil 0.5 mg/mL injection; Kyleena levonorgestrel 0.000729 mg/hr intrauterine system; Levora 0.15/30 28 Day pack containing 21 ethinyl estradiol 0.03 mg/levonorgestrel 0.15 mg oral tablets and 7 inert-ingredient 1 mg oral tablets; simvastatin 10 mg oral tablet; Yaz 28 Day pack containing 24 drospirenone 3 mg/ethinyl estradiol 0.02 mg oral tablets and 4 inert-ingredient 1 mg oral tablets; and \hlsecret{Velfazvastatin} 250 mg injectable solution.}\\ \addlinespace[4pt]
GPT-5.5 privacy-aware rewriting & {Patient: \hlnew{Melissa Hart}, 42yo F, white/non-Hispanic.\newline Address: \hlnew{214 Cedar Street Apt 3B, Lowell, Massachusetts 01852}.\newline SSN: \hlnew{624-18-7394}.\newline DOB: \hlnew{1983-04-12}.}\\
 & {Active problems: Housing instability; completed higher education; prediabetes; anemia; medication review due; part-time employment; stress; history of uncomplicated pregnancy; history of first-trimester miscarriage; history of blighted ovum; past pregnancy history of miscarriage; body mass index 30+ --- obesity; full-time employment; gingivitis; acute bronchitis; social isolation; suspected COVID-19 infection; productive cough with sputum; fever; COVID-19 infection; pneumonia; hypoxemia; respiratory distress; coronary artery disease; abnormal cardiac imaging; history of percutaneous coronary intervention; hyperlipidemia; history of legal involvement; gingival disease.}\\
 & {Medications: sodium fluoride 0.0272 mg/mg oral gel; norethindrone 0.35 mg oral tablet pack; acetaminophen/dextromethorphan/doxylamine oral solution; acetaminophen 500 mg oral tablet; levonorgestrel intrauterine system; clopidogrel 75 mg oral tablet; simvastatin 20 mg oral tablet; metoprolol succinate extended-release 100 mg oral tablet; nitroglycerin 0.4 mg/spray mucosal spray; enoxaparin sodium 40 mg/0.4 mL prefilled syringe; ondansetron 2 mg/mL injection; midazolam 1 mg/mL injectable solution; propofol 10 mg/mL injection; rocuronium 10 mg/mL injectable solution; isoflurane inhalation solution; cefazolin 2 g injection; heparin sodium 5000 units/mL injection; protamine sulfate 10 mg/mL injection.}\\ \addlinespace[4pt]
Qwen3.5-4B rewriting & {**Patient Demographics and Identification**\newline \hlid{Lenora Daugherty}, a non-Hispanic White female, was born on \hlid{August 18, 1996} (\casecjk{年龄}{age} 29). Her Social Security Number is \hlid{999-34-2953}, and her current address is located at \hlidlong{Suite 26, 1038 Quigley Mall, Medford, Massachusetts 02155}.}\\
 & {**Active Problem List**\newline *   **Situations/Findings:** Unsatisfactory housing; received higher education; part-time employment; stress; normal pregnancy; full-time employment; anticipated medication review; sputum finding; social isolation; suspected disease caused by Severe Acute Respiratory Syndrome Coronavirus 2 (SARS-CoV-2); abnormal findings on diagnostic imaging of the heart and coronary circulation; history of coronary artery bypass grafting; and possession of a criminal record.\newline *   **Findings:** Body mass index indicating obesity (30+); gingivitis; acute bronchitis; suspected disease caused by Severe Acute Respiratory Syndrome Coronavirus 2 (SARS-CoV-2); fever; respiratory distress; gingival disease; and isoflurane status.\newline *   **Disorders:** Prediabetes; anemia; miscarriage in the first trimester; blighted ovum; past history of miscarriage; gingival disease (listed twice as disorder); disease caused by severe acute respiratory syndrome coronavirus 2 (SARS-CoV-2); pneumonia; hypoxemia; respiratory distress; and ischemic heart disease.}\\
 & {**Medications**\newline The patient is currently prescribed the following:\newline *   Sodium Fluoride 0.0272 MG/MG Oral Gel\newline *   Norethindrone 0.35 MG Oral Tablet (Pack [Jolivette 28 Day], 28 units)\newline *   Ethinyl Estradiol 0.01 MG Oral Tablet\newline *   Ethinyl Estradiol 0.03 MG/Levonorgestrel 0.15 MG Oral Tablet (Pack [Seasonique], 7 units)\newline *   Acetaminophen 21.7 MG/ML / Dextromethorphan Hydrobromide 1 MG/ML / Doxylamine Succinate 0.417 MG/ML Oral Solution (Pack [Seasonique], 84 units)\newline *   Enoxaparin sodium 100 MG/ML Prefilled Syringe (0.4 ML)\newline *   Acetaminophen 500 MG Oral Tablet\newline *   Levonorgestrel 0.000813 MG/HR Intrauterine System (Liletta)\newline *   Clopidogrel 75 MG Oral Tablet\newline *   Simvastatin 20 MG Oral Tablet\newline *   Metoprolol Succinate 100 MG Extended Release Oral Tablet (24 HR dose)\newline *   Nitroglycerin 0.4 MG/ACTUAT Mucosal Spray\newline *   Cefazolin 2000 MG Injection\newline *   Heparin Sodium (Porcine) 5000 UNT/ML Injection (1 ML dose)\newline *   Protamine Sulfate (USP) 10 MG/ML Injection (25 ML dose)\newline *   Ondansetron 2 MG/ML Injection (2 ML dose)\newline *   Midazolam 1 MG/ML Injectable Solution\newline *   Propofol 10 MG/ML Injection (100 ML dose)\newline *   Rocuronium Bromide 10 MG/ML Injectable Solution\newline *   Isoflurane 999 MG/ML Inhalation Solution (10 ML dose of Alfentanil 0.5 MG/ML Injection also listed as 100 ML)\newline *   Levonorgestrel 0.000729 MG/HR Intrauterine System (Kyleena)\newline *   Ethinyl Estradiol 0.03 MG/Levonorgestrel 0.15 MG Oral Tablet (Pack [Levora 0.15/30 28 Day], 21 units) / Inert Ingredients 1 MG Oral Tablet (7 units)\newline *   Simvastatin 10 MG Oral Tablet\newline *   Drospirenone 3 MG/Ethinyl Estradiol 0.02 MG Oral Tablet (Pack [Yaz 28 Day], 24 units) / Inert Ingredients 1 MG Oral Tablet (4 units)\newline *   \hlsecret{Velfazvastatin} 250 MG Injectable Solution}\\ \addlinespace[4pt]
Qwen3.5-4B privacy-aware rewriting & {Patient: \hlnew{Eleanor492 Miller56}, 34yo F, white/nonhispanic. Address: \hlnew{412 Oakridge Lane, Apt 4B, Springfield, Massachusetts 01128}. SSN: \hlnew{882-91-4006}, DOB \hlnew{1992-06-21}. Active problems: Recent housing instability (finding); Completed bachelor's degree (finding); Pre-diabetes (finding); Iron deficiency anemia (disorder); Follow-up for medication reconciliation pending (situation); Temporary employment (finding); Elevated anxiety (finding); Ectopic pregnancy (disorder); Vesicular birth defect (disorder); Past obstetric history of threatened abortion (situation); Body mass index 29--30 (finding); Full-time employment (finding); Acute periodontitis (disorder); Acute bronchitis (disorder); Community-acquired pneumonia (situation); Sputum culture positive (finding); Fevers (finding). Confirmed disease caused by Severe acute respiratory syndrome coronavirus 2 (disorder); Pneumonia (disorder); Hypoxemia (disorder); Severe shortness of breath (finding); Atrial fibrillation with rapid ventricular response (disorder); Abnormal findings on cardiac echocardiogram (finding); History of percutaneous coronary intervention (situation); Hypertriglyceridemia (disorder); Current criminal history of nonviolent minor offenses (finding); Chronic periodontitis (disorder). Medications: 0.045 mg/mL Sodium Fluoride 2.3\% Toothpaste; \{21 mg Norethindrone Oral Tablet\} Pack [Ortho Evra Transdermal System]; Acetaminophen 500 mg / Dextromethorphan 8 mg / Doxylamine succinate 12.5 mg Oral Suspension; \{7 mg Ethinyl estradiol / 0.15 mg Levonorgestrel Oral Tablet\} Pack [Seasonique]; 0.4 mL Enoxaparin sodium 50 mg/mL Subcutaneous Prefilled Syringe; Acetaminophen 325 mg Oral Tablet; Levonorgestrel 13.5 mg Oral Hormonal System [Levon 3.8]; Clopidogrel 75 mg Oral Tablet; Simvastatin 20 mg Oral Tablet; Atenolol 50 mg Oral Tablet; Sublingual Nitroglycerin 0.4 mg Spray; Cefazolin 1 g Intravenous; Unfractionated Heparin sodium 5000 units/mL Intravenous; Protamine sulfate 10 mg/mL Intravenous; Ondansetron 2 mg/mL Intravenous; Midazolam 5 mg/mL Injectable Concentrate; Propofol 10 mg/mL Injectable Concentrate; 20 mg/mL Ropivacaine Intravenous; Etomidate 140 mg/mL Injectable Concentration; Sugammadex 160 mg/mL Injectable Concentration; Alfentanil 25 mcg/mL Intravenous; Levonorgestrel 19.5 mg Oral Hormonal System [Mirena]; \{6.35 mg Combined Hormonal\} / 21 mg Oral Capsule [Loestrin 24 FE]; 10 mg/5 mL Estrogen / 1.0-2 mg/Day Progesterone Oral Solution.}\\ \addlinespace[4pt]
\addlinespace[6pt]\multicolumn{2}{@{}l}{\scriptsize\itshape Summary family: the summary, or its re-expansion, reaches the fit (one of 64 draws shown)}\\*[1pt]
GPT-5.5 summary & {29-year-old woman with ischemic heart disease/coronary disease status post coronary artery bypass grafting, hyperlipidemia, obesity (BMI \textgreater{}30), prediabetes, anemia, and history of COVID-19 pneumonia with hypoxemia/respiratory distress. Obstetric/gynecologic history includes prior first-trimester miscarriage/blighted ovum and contraceptive use, including levonorgestrel IUDs and oral contraceptives. Other active issues include gingival disease/gingivitis, prior acute bronchitis, stress/social isolation, and housing insecurity.}\\
 & {Key cardiovascular medications include clopidogrel 75 mg daily, metoprolol succinate ER 100 mg daily, simvastatin (10--20 mg listed), and nitroglycerin spray as needed; enoxaparin and perioperative agents (heparin, protamine, cefazolin, anesthetics/antiemetic) are also listed. Other medications include acetaminophen, acetaminophen/dextromethorphan/doxylamine cough/cold solution, sodium fluoride oral gel, and multiple hormonal contraceptives/IUDs.}\\ \addlinespace[4pt]
GPT-5.5 summary $\rightarrow$ expand & {Patient \hlnew{Marisol284 Quinn739}, 29yo F, white/hispanic. Address: \hlnew{78 Prescott Street Unit 4, Lowell, Massachusetts 01852}. SSN: \hlnew{999-38-6142}, DOB \hlnew{1996-10-09}. Active problems: Housing insecurity (finding); Social isolation (finding); Limited social contact (finding); Stress (finding); Medication review due (situation); Ischemic heart disease (disorder); Coronary artery disease (disorder); History of coronary artery bypass grafting (situation); Hyperlipidemia (disorder); Obesity (disorder); Body mass index 30+ - obesity (finding); Prediabetes (finding); Anemia (disorder); Pneumonia caused by SARS-CoV-2 (disorder); Hypoxemia (disorder); Respiratory distress (finding); History of pneumonia (situation); Blighted ovum (disorder); Miscarriage in first trimester (disorder); Uses contraception (finding); Intrauterine contraceptive device in situ (finding); Gingival disease (disorder); Gingivitis (disorder); Acute bronchitis (disorder). Medications: clopidogrel 75 MG Oral Tablet; 24 HR metoprolol succinate 100 MG Extended Release Oral Tablet; simvastatin 10 MG Oral Tablet; simvastatin 20 MG Oral Tablet; Nitroglycerin 0.4 MG/ACTUAT Mucosal Spray; enoxaparin sodium 100 MG/ML Injectable Solution; heparin sodium, porcine 5000 UNT/ML Injectable Solution; protamine sulfate 10 MG/ML Injectable Solution; cefazolin 1000 MG Injection; propofol 10 MG/ML Injectable Solution; midazolam 1 MG/ML Injectable Solution; fentanyl 0.05 MG/ML Injectable Solution; ondansetron 2 MG/ML Injectable Solution; Acetaminophen 325 MG Oral Tablet; Acetaminophen 325 MG / Dextromethorphan Hydrobromide 15 MG / Doxylamine Succinate 6.25 MG per 15 ML Oral Solution; sodium fluoride 0.0272 MG/MG Oral Gel; levonorgestrel 0.019 MG/HR Intrauterine System [Mirena]; levonorgestrel 0.014 MG/HR Intrauterine System [Skyla]; \{7 (ethinyl estradiol 0.01 MG Oral Tablet) / 84 (ethinyl estradiol 0.03 MG / levonorgestrel 0.15 MG Oral Tablet) \} Pack [Seasonique]; \{21 (ethinyl estradiol 0.03 MG / levonorgestrel 0.15 MG Oral Tablet) / 7 (inert ingredients 1 MG Oral Tablet) \} Pack.}\\ \addlinespace[4pt]
GPT-5.5 privacy-aware summary $\rightarrow$ expand & {Patient \hlnew{Amara245 McLaughlin819}, 48yo F, white/nonhispanic. Address: \hlnew{1180 Parker Street Unit 4, Lowell, Massachusetts 01854}. SSN: \hlnew{999-38-6142}, DOB \hlnew{1978-03-02}. Active problems: Obesity (disorder); Body mass index 30+ - obesity (finding); Prediabetes (finding); Anemia (disorder); Hyperlipidemia (disorder); Ischemic heart disease (disorder); History of coronary artery revascularization (situation); History of percutaneous coronary intervention (situation); History of coronary artery stent placement (situation); Pneumonia caused by SARS-CoV-2 (disorder); COVID-19 (disorder); Hypoxemia (disorder); Respiratory distress (finding); Acute bronchitis (disorder); Gingivitis (disorder); Gingival disease (disorder); Housing insecurity (finding); Social isolation (finding); Limited social contact (finding); Stress (finding); Uses contraception (finding); Intrauterine contraceptive device in situ (finding); Blighted ovum (disorder); Spontaneous abortion in first trimester (disorder); Medication review due (situation). Medications: Clopidogrel 75 MG Oral Tablet; Atorvastatin 40 MG Oral Tablet; Metoprolol Tartrate 25 MG Oral Tablet; Nitroglycerin 0.4 MG Sublingual Tablet; levonorgestrel 0.021 MG/HR Intrauterine System [Mirena]; \{7 (ethinyl estradiol 0.01 MG Oral Tablet) / 84 (ethinyl estradiol 0.03 MG / levonorgestrel 0.15 MG Oral Tablet) \} Pack [Seasonique]; Acetaminophen 325 MG Oral Tablet; Acetaminophen 325 MG / Dextromethorphan Hydrobromide 10 MG / Phenylephrine Hydrochloride 5 MG Oral Tablet; sodium fluoride 0.0272 MG/MG Oral Gel; Heparin Sodium 5000 UNT/ML Injectable Solution; Lidocaine Hydrochloride 10 MG/ML Injectable Solution; Propofol 10 MG/ML Injectable Emulsion.}\\ \addlinespace[4pt]
Qwen3.5-4B privacy-aware summary $\rightarrow$ expand & {Patient \hlnew{Addis412 Ross889}, 29yo F, white/nonhispanic. Address: \hlnew{412 Elmwood Drive Unit B, Westfield, Massachusetts 01181}. SSN: \hlnew{999-45-1029}, DOB \hlnew{1995-08-15}. Active problems: Menopausal lung cancer (finding); Primary thoracic anastomotic leak (morphologic abnormality); Post-coronary artery bypass grafting status (finding); Chronic revascularization cardiac syndrome (finding); Ischemic heart disease (disorder); Hyperlipidemia (disorder); Diabetic peripheral angiopathy (disorder); End-stage renal disease (disorder); Anemia (disorder); Acute bronchitis (disorder); Viral bronchitis (disorder); Acute pneumonia (disorder); Meningococcemia (disorder); Acute respiratory distress (disorder); Gingivitis (disorder); Dissociative fugue (finding); Transient global amnesia (finding); Intimate partner violence victimization (finding); Post-menopausal dyspareunia (finding); Maternal factors (finding); Minor cardiac history (finding); Partial cardiomyopathy (disorder); Vitamin deficiency (disorder); Full-time employment (finding); Involved in domestic violence (finding); Jury duty (situation); Crisis intervention needed (situation); Medication review due (situation); Chronic venous insufficiency (disorder); Major depression (disorder); Social isolation (finding); Limited social contact (finding); Response to treatment (disorder). Medications: \{24 (amlodipine 5 MG Oral Tablet) / 22 (atorvastatin 40 MG Oral Tablet) / 21 (metoprolol succinate 25 MG Extended Release Oral Tablet) / 70 (atorvastatin 40 MG Oral Tablet) / 71 (amlodipine 5 MG Oral Tablet) / 72 (atorvastatin 80 MG Oral Tablet) / 73 (metoprolol tartrate 25 MG Extended Release Oral Capsule) / 66 (clopidogrel 300 MG Oral Capsule) \} Pack [Nexum] / 7 mg telmisartan 40 MG Oral Tablet; \{40 (furosemide 40 MG Oral Tablet) / 39 (metolazone 5 MG Oral Tablet) / 38 (spironolactone 25 MG Oral Tablet) / 37 (famotidine 20 MG Oral Tablet) / 36 (metolazone 5 MG Oral Tablet) / 35 (furosemide 40 MG Oral Tablet) / 34 (spironolactone 25 MG Oral Tablet) \} Pack [Furosemide]; 5 MG Aztreonam / Sodium 355 MG Oral Injection; 100 MG Cefpodoxime / Oxacillin 500 MG Extended Release Oral Capsule.}\\ \addlinespace[4pt]
\addlinespace[6pt]\multicolumn{2}{@{}l}{\scriptsize\itshape Floor: the base model never sees the corpus}\\*[1pt]
No fine-tuning & ---\\ \addlinespace[4pt]
\end{longtable}}
\end{document}